\documentclass[11pt]{article}
\usepackage{amsmath}
\usepackage{amssymb}
\usepackage{amsthm}
\usepackage{latexsym}
\usepackage{color}
\usepackage{graphics}
  \usepackage{graphicx}
\usepackage{appendix}
\usepackage{color}
\usepackage{epstopdf}
\usepackage{relsize}
\usepackage{enumerate}
\usepackage[T1]{fontenc}
\usepackage[english]{babel}
\usepackage[table]{xcolor}
\usepackage{algorithm, algorithmic}
\usepackage{amsmath, amsfonts}
\usepackage{multirow}
\DeclareSymbolFont{calletters}{OMS}{cmsy}{m}{n}
\DeclareSymbolFontAlphabet{\mathcal}{calletters}
\def\be{\begin{eqnarray}}
\def\ee{\end{eqnarray}}

\def\b*{\begin{eqnarray*}}
\def\e*{\end{eqnarray*}}

\makeatletter \@addtoreset{equation}{section}

\@addtoreset{Definition}{section}

\@addtoreset{Theorem}{section}

\@addtoreset{Proposition}{section}

\@addtoreset{Property}{section}

\@addtoreset{Assumption}{section}

\@addtoreset{Corollary}{section}

\@addtoreset{Lemma}{section}

\@addtoreset{Remark}{section}

\@addtoreset{Example}{section}

\@addtoreset{Condition}{section}

\def \N{\mathbb{N}}

\def \R{\mathbb{R}}

\def\Lc{{\cal L}}

\def\Nc{{\cal N}}

\def\Sc{{\cal S}}

\def\XX{{\bf X}}

\def\={\;=\;}
\def\.{\;.}

\def\1{{\bf 1}}

\def\b*{\begin{eqnarray*}}
\def\e*{\end{eqnarray*}}

 \def\normeL2#1{\left\|{#1}\right\|_{L^2}}

\title{The behavior of dealers and clients on the European corporate bond market: the case of Multi-Dealer-to-Client platforms\thanks{This research has been conducted with the support of the Research
Initiative ``Nouveaux traitements pour les donn\'ees lacunaires issues des activit\'es de cr\'edit'' financed by BNP Paribas under the aegis of the Europlace Institute of Finance. The content of this article does not reflect in any manner the practices of BNP Paribas.}}

\author{Jean-David Fermanian\footnote{ENSAE-CREST, 3 avenue Pierre Larousse, 92245 Malakoff Cedex, France.},
Olivier Gu\'eant\footnote{ENSAE-CREST, 3 avenue Pierre Larousse, 92245 Malakoff Cedex, France.}, Jiang Pu\footnote{LPMA, Universit\'e Paris-Diderot, avenue de France, 75205 Paris Cedex 13, France. Institut Europlace de Finance, 28 place de la Bourse, 75002 Paris, France.}}

\date{}

\begin{document}

\maketitle

\begin{center}
\textbf{Abstract}
\end{center}

For the last two decades, most financial markets have undergone an evolution toward electronification. The market for corporate bonds is one of the last major financial markets to follow this unavoidable path. Traditionally quote-driven (\emph{i.e.}, dealer-driven) rather than order-driven, the market for corporate bonds is still mainly dominated by voice trading, but a lot of electronic platforms have emerged. These electronic platforms make it possible for buy-side agents to simultaneously request several dealers for quotes, or even directly trade with other buy-siders. The research presented in this article is based on a large proprietary database of requests for quotes (RFQ) sent, through the multi-dealer-to-client (MD2C) platform operated by Bloomberg Fixed Income Trading, to one of the major liquidity providers in European corporate bonds. Our goal is (i) to model the RFQ process on these platforms and the resulting competition between dealers, and (ii) to use our model in order to implicit from the RFQ database the behavior of both dealers and clients on MD2C platforms.

\section{Introduction}

For many years, the trading of corporate bonds\footnote{In this article, the term ``corporate bonds'' encompasses most non-government bonds, issued by financial institutions or other corporates.} on the secondary market only took place over the counter via private negotiations on the phone. The organization of the market corresponded to a classical quote-driven one, where market participants are divided into two groups: clients (the buy side), \emph{i.e.}, asset management companies, pension funds, insurance companies, wealth management companies, and hedge funds, who call dealers (the sell side) -- mainly investment banks -- for buying or selling bonds. In this economic environment, dealers act as market makers and liquidity providers, whereas clients are liquidity takers. The market was also divided into two segments: the dealer-to-client (D2C) segment, where transactions occur between dealers and clients, and the interdealer-broker (IDB) segment.\\

This traditional description of the corporate bond market is still valid in many ways, but the way corporate bonds are traded is evolving constantly. Recent evolutions are due to at least two factors: technological innovation and financial regulation.\\

The electronification of financial markets has dramatically changed the way most securities are traded, and subsequently the organization of most financial markets. Electronic order books are now the norm for a lot of asset classes: stocks obviously, but also U.S. Treasuries and foreign exchanges. Although many historically dealer-driven markets eventually adopted the order-driven paradigm,\footnote{In the case of the UK, the transition occurred in 1997, with the creation of the SETS.} it is difficult to imagine a similar market organization for corporate bonds today. One major difference between bonds and stocks, to remain in the field of cash markets, has to do with heterogeneity and liquidity. There is usually one stock for a given company while there are often dozens of bonds for the same company, corresponding to different maturities, different coupons, and different seniorities. A natural consequence is that most corporate bonds are illiquid. The Securities Industry and Financial Markets Association (SIFMA) estimates indeed (see \cite{scottquinn}) that the total market value of stocks is twice the market value of corporate bonds, but that there are more than six times more listed corporate bonds than listed stocks, and furthermore, the global average daily volume is estimated to be \$17.9 billion for corporate bonds vs. \$112.9 billion for stocks. Even worse, MarketAxess Research (see \cite{mck}) estimates that, in 2012, 38\% of the 37,000 TRACE-eligible bonds did not trade, even once, and only 1\% of these 37,000 bonds traded every day.
The lack of standardization and the illiquidity of most corporate bonds, along with the buy-and-hold strategy of many investors, make almost impossible the emergence of a match-based model with order books similar to those of equity markets, except perhaps for the most liquid corporate bonds. Nevertheless, electronification occurs, and new models emerge that are different from those of equity markets.\\

Technological innovation occurs in the D2C market segment which is ``electronifying'' at a rapid pace, whereas trades in the IDB market segment remain almost entirely executed via voice.\footnote{Although it is still high, the volume of the IDB market segment has been constantly decreasing for years. The IDB segment is not the purpose of this paper.} Several types of platforms have emerged. The most important and successful ones are multi-dealer-to-client (MD2C) trading platforms. Examples of such MD2C platforms are those proposed by Bloomberg Fixed Income Trading (FIT), Tradeweb and MarketAxess. They clearly dominate the landscape of electronic trading. With these platforms, there is still a distinction between dealers and clients, but clients can send simultaneously a request for quotes to several dealers who have streamed prices to the platforms. Another kind of MD2C electronic platform that is used by many buy-siders (especially small ones) consists of executable quotes, but only for odd lots. Single-dealer electronic platforms have also emerged to replace the telephone. Crossing systems, and platforms proposing all-to-all central limit order books (CLOB) also exist. They try to revolutionized the classical distinction between dealers who provide liquidity and clients who take liquidity, but they are still very rarely used in practice.\\

An interesting study carried out in \cite{mck} shows that the current market structure is far from being stabilized, given the different viewpoints of market participants on the future organization of the market. If it is commonly believed that the market will remain for years a dealer-driven one, with MD2C RFQ platforms holding the lion's share of electronic trading, the evolution of financial regulation will have a lot of influence on the market structure and the role of the different participants. It is already clear that Basel III capital requirement deters investment banks from holding large inventories. Therefore, their traditional role as dealers may change from being market makers to simply providing access to the market. This might encourage all-to-all platforms and give a new role to large asset managers as specialists on these platforms. In addition to Basel III, the question of pre-trade transparency for corporate bonds has been raised in the debate on MiFID 2: the European landscape may change to apply new rules.\\

In this paper, we focus on the current state of the market, and on the dominant MD2C platform (Bloomberg FIT), where buy-siders send quotes to a swath of dealers on a specific bond. More precisely, the process works as follows in the case of a client who wants to buy/sell a given bond:\footnote{Sometimes, clients send RFQs with no intention to buy or sell bonds, but only to get information.}

\begin{enumerate}
  \item The client connects to the platform and sees the bid and offer prices streamed by the dealers for the security. These prices, streamed by dealers, correspond to prices for a given reference size. These streamed prices are not firm/binding prices.
  \item The client selects dealers (up to 6 dealers on Bloomberg FIT), and sends one RFQ through the platform to these dealers, with a precise volume (notional), and the side (``buy'' order or ``sell'' order).
  \item Requested dealers can answer a price to the client for the transaction (not necessarily the price he/she has streamed). Dealers know the identity of the client (contrary to what happens in the case of most of the systems organized around a CLOB) and the number of requested dealers (the degree of competition for this request). However, they do not see the prices that have been streamed by other dealers. They only see a composite price at the bid and offer, based on some of the best streamed prices. These composite prices in the case of Bloomberg FIT are called the CBBT bid price and the CBBT offer price.
  \item The client progressively receives the answers to the RFQ. He/She can deal at any time with the dealer who has proposed the best price, or decide not to trade.
  \item Each dealer (who has answered a price) knows whether a deal was done (with him/her, but also with another dealer -- without knowing the identity of that dealer) or not.
  If a transaction occurred, the best dealer usually knows the cover price, if there is one.
  \end{enumerate}

The electronification of the request process between supply-siders and liquidity providers generates a lot of data. For each RFQ they receive, dealers can record information. Therefore, dealers now have a complete history of their interactions with their clients, in a very standardized fashion, and they even get notified of trades occurring
with competitors when they participate in a RFQ process. \\

The work presented in this article is based on a database of RFQs received by BNP Paribas -- one of the most important dealers in European corporate bonds --  through Bloomberg FIT. BNP Paribas  will be designated throughout this paper by the term ``reference dealer'', even though it does not play any particular role in the market. The proprietary database we got access to represents a fraction of the RFQs received by the reference dealer over nearly two years (2014-2015). For each RFQ, we observe its characteristics (date, hour, id of the client, Isin of the bond, buy/sell side, notional, number of dealers requested, etc.), contextual information (prices streamed and answered by the reference dealer, CBBT bid and offer prices, etc.), and the outcomes of the RFQ (whether or not there was a deal, the cover price -- if there is one -- in the case of a trade with the reference dealer, etc.).\\

We build a parsimonious model for the RFQ process, in which the (unobserved) prices answered by the dealers competing with the reference dealer follow an unknown distribution, and where clients decide to trade or not to trade depending on a (unobserved) reservation value following another unknown distribution. By using log-likelihood maximization techniques, we estimate these distributions for the quotes answered by the dealers and for the reservation value of the clients. These distributions are parameterized to take account of the information available. Therefore, we address questions such as the dependence of the behavior of dealers on the number of dealers requested, the notional, etc. Similarly, we analyze how buy-siders behave depending on the context.\\

Applications of our work are numerous. Assessing the behavior of competitors is indeed of the utmost importance for a dealer. By modelling the behavior of his competitors in a better way, a dealer can expect to better analyze/manage hit ratios, and better estimate the probability to trade at a given price. Competition can also be analyzed: the behavior of dealers answering a RFQ indeed depends on the degree of competition, \emph{i.e.}, the number of dealers requested. A good model to estimate the reservation value of clients with respect to a bond is also key for a dealer: it is one of the most important inputs for choosing the quote he/she answers to the client in the RFQ process. Our model leads to very general input functions for market making models. In a nutshell, the goal of market makers / dealers, is to quote bid and offer prices so as to make money out of the spread between these two prices, while mitigating the risk of price changes on the value of the inventory. Old market making models include those of Ho and Stoll \cite{hs1,hs2}. More recently, Avellaneda and Stoikov \cite{as} (see also \cite{gueant2013dealing} for closed-form expressions for the optimal quotes) proposed a model that can be applied to quote-driven markets such as the corporate bond market.\footnote{The model was initially built for market making on stock markets, but it is better suited for market making on quote-driven markets.} However, the exponential form of the execution intensities used in \cite{as} and \cite{gueant2013dealing} is questionable in practice. The more general models presented in the appendix of Gu\'eant and Lehalle \cite{gueant2013general}, in \cite{gueantlivre} and in \cite{gueantmm} go beyond the case of exponential intensities, and they may be adapted to incorporate the findings of this paper in a dynamic market making model.\\

In Section 2, we describe our model for the RFQ process. In Section 3, we present the features of our dataset and our estimation method. Empirical results are shown in Section~4.

\section{The model for dealers' quotes and clients' behavior}
\label{The_model}

\subsection{Description of the variables}

We have described above the RFQ process on the Bloomberg FIT platform. Now, let us specify the model we propose for the behavior of dealers and clients, along with the associated notations.\\

Note that we adopt the point of view of the so-called ``reference dealer'' throughout this paper. In particular, we propose a model to understand the behavior of other dealers (competitors) and clients during RFQs answered by the reference dealer. We assume that the reference dealer decision-making process for answering a RFQ is independent of the quoting process of other dealers. Furthermore, we assume that the behavior of clients does not depend on whether or not the reference dealer is requested. Implicitly, some irrational behaviors, as those of ``captive clients'' who would accept any quote proposed by the reference dealer, are assumed not to exist.\\

Let us consider first the case of a RFQ $i$ (answered by the reference dealer) corresponding to a ``buy'' order.\footnote{A ``buy'' (resp. ``sell'') order means that the client sends a request to buy (resp. sell) the bond. The dealer then answers a quote to sell (resp. buy) the bond.} Our model works as follows:

\begin{enumerate}
\item A (prospective) client has identified a specific corporate bond that may be of interest for him/her. He/she sees the prices streamed by dealers for this bond, and believes that it is worth sending a (buy) RFQ.\footnote{The client may believe that the value of this bond is likely to increase in the future, and he/she expects to sell it back. Alternatively, he may think that a buy-and-hold strategy is profitable, given the current evaluation of the market for that corporate bond.} We denote by $V_i$ the client's view on the value of the bond.
\item This client sends a RFQ to $n_i+1$ dealers -- the reference dealer plus $n_i \in \{0,\ldots,5\}$ other dealers -- to buy a number of bonds corresponding to a specific amount of cash (the notional of the RFQ). Each dealer sees the CBBT bid and offer prices which are composite prices based on the prices streamed by dealers. In addition to the reference dealer, whose answered price is denoted by $Y_i$, each dealer can answer a price to the client. In the first version of our model, we assume that the dealers always answer, and we denote by $W_{k,i}$ the answer of dealer $k$ ($k=1,\ldots,n_i$). In the second version, we assume that each dealer may answer or not answer, and we denote by $A_{k,i}\in \{0,1\}$ the occurrence of an answer for the dealer $k$. The prices $Y_i$ and $W_{k,i}$ are binding, in the sense that $Y_i$ (resp. $W_{k,i}$) will be the transaction price if the reference dealer (resp. the dealer $k$) is chosen by the client for this deal.
\item We assume that a deal occurs if and only if a dealer proposes a price lower than $V_i$. For that reason, we refer to $V_i$ as the reservation price or reservation value of the client. In that case, the transaction occurs between the client and a dealer chosen by the client amongst the dealers who have proposed the lowest price. In particular, the deal price is $\min(Y_i,\min_{k=1,\ldots,n_i} W_{k,i})$ in our first model and $\min(Y_i,\min_{\lbrace k=1,\ldots,n_i | A_{k,i} = 1\rbrace} W_{k,i})$ in the second model.\\
\end{enumerate}

The above mechanism applies to a ``buy'' order, but a symmetrical one applies in the case of a ``sell'' order. In that case indeed, a deal occurs if and only if a dealer proposes a price greater than the client's reservation price $V_i$, and the deal price is then  $\max(Y_i,\max_{k=1,\ldots,n_i} W_{k,i})$ in our first model and $\max(Y_i,\max_{\lbrace k=1,\ldots,n_i | A_{k,i} = 1\rbrace} W_{k,i})$ in the second model.\\

Obviously, the whole previous process depends on the bond characteristics and the public market information at the time of the RFQ. We assume that this  information is summarized into a $\sigma$-algebra $\Omega_i$.\\

\subsection{Distributional assumptions}

We model both the reservation value $V_i$ of clients and the price answered by dealers $W_{k,i}$ by random variables. Throughout this paper, $F$ (or in fact $F(\cdot\left|\Omega_i)\right.$) refers to the cumulative distribution function of the variables $W_{k,i}$, and $G$ (or in fact $G(\cdot\left|\Omega_i)\right.$) refers to the cumulative distribution function of the variable $V_{i}$, in the case of a ``buy'' RFQ. We denote by $f$ and $g$ the corresponding probability density functions.
In the case of a ``sell'' RFQ, the same notations are used, with a star: $F^*$, $G^*$, etc. It is noteworthy that the distributions of dealers' quotes are assumed to be the
same across the dealers in competition with the reference dealer (but not necessarily equal to the distribution of the reference dealer's quotes). Moreover, for the sake of simplicity, we assume this is the case for clients too, even though it is possible -- and easy -- to associate different functions $G$ and $G^*$ to different clients.\\

To complete the model specifications, we need to state our assumptions concerning the functional form of the functions $F$ and $G$, and similarly for $F^*$ and $G^*$.\\

Because the CBBT mid-price constitutes a reference price, and because the CBBT bid-to-mid $\Delta_i$ (\emph{i.e.}, half the CBBT bid-ask spread\footnote{One could alternatively use the streamed bid-ask spread.}) constitutes a proxy of liquidity -- which is crucially linked to the level of risk aversion associated with bond quoting --, it is convenient to work with ``reduced quotes'': $(V_i-\text{CBBT}_i)/\Delta_i$ for clients, and $(W_{k,i}-\text{CBBT}_i)/\Delta_i$ for dealers, where $\text{CBBT}_i$ is the CBBT mid-price.\\

In other words, it makes sense to assume that
\begin{equation}
\label{specif1}
 F( \xi|\Omega_i ) = F_0\left(\frac{\xi-\text{CBBT}_i}{\Delta_i} \right), \qquad F^*( \xi|\Omega_i ) = F^*_0\left(\frac{\xi-\text{CBBT}_i}{\Delta_i} \right),
\end{equation}
\begin{equation}
\label{specif2}
 G( \xi|\Omega_i ) = G_0\left(\frac{\xi-\text{CBBT}_i}{\Delta_i} \right), \qquad G^*( \xi|\Omega_i ) = G^*_0\left(\frac{\xi-\text{CBBT}_i}{\Delta_i} \right),
\end{equation}
for some cumulative distribution functions $F_0$, $F^*_0$, $G_0$, and  $G^*_0$.\footnote{CBBT prices are included into $\Omega_i$.}\\

In (\ref{specif1}) and (\ref{specif2}), there is no dependency on the number of dealers. In what follows, we often consider instead the following specification:
\begin{equation}
\label{Fn}
 F( \xi|\Omega_i )
=F_0\left(\frac{\xi-\text{CBBT}_i}{\Delta_i} ; n_i \right), \qquad F^*( \xi|\Omega_i ) = F^*_0\left(\frac{\xi-\text{CBBT}_i}{\Delta_i} ;n_i \right),
\end{equation}
\begin{equation}
\label{Gn}
 G( \xi|\Omega_i ) = G_0\left(\frac{\xi-\text{CBBT}_i}{\Delta_i} ;n_i \right), \qquad G^*( \xi|\Omega_i ) = G^*_0\left(\frac{\xi-\text{CBBT}_i}{\Delta_i} ; n_i\right),
\end{equation}
for some cumulative distribution functions $F_0(\cdot;n)$, $F_0^*(\cdot;n)$, $G_0(\cdot;n)$, and $G_0^*(\cdot;n)$, where $n\in \{0,\ldots,5\}$ corresponds to the number of dealers requested in addition to the reference dealer.\\

In this paper, we rely on a parametric specification of the above distributions. In the large database of RFQs used to carry out the research presented in this paper, we have observed that the empirical distribution of $\left((Y_i - \text{CBBT}_i )/\Delta_i \right)_i$ is leptokurtic (fat tailed),
very spiky around the composite price and asymmetric. Assuming most dealers should behave in a similar manner, we have decided to use the skew exponential power (SEP) distribution, for which the skewness and the kurtosis have been shown to belong to rather wide intervals.\\

As a preliminary, let us recall the exponential power distribution (see \cite{subotin}) characterized by the probability density function
$$
 f_{EP}(x;\mu,\sigma,\alpha)= \frac{1}{c\sigma}\exp\left(  -|z|^\alpha /\alpha \right), x\in \R,$$
where $\alpha>0$, $\mu \in \R$, $\sigma>0$, $z:=(x-\mu)/\sigma$ and $c:= 2\alpha^{1/\alpha -1} \Gamma (1/\alpha)$.\\

The density of the SEP distribution is deduced from the general methodology to accommodate asymmetry proposed by Azzalini in \cite{Azzalini85}: one introduces an additional parameter $\lambda \in \R$ that reflects asymmetry, and the density of the SEP distribution is given by
\begin{equation}
f_{SEP} (x)= 2\Phi\left(w\right)f_{EP}(x;\mu,\sigma,\alpha),\; \; x\in \R,
\end{equation}
where $w:=\text{sign}(z) |z|^{\alpha/2} \lambda (2/\alpha)^{1/2}$, and $\Phi$ is the cumulative distribution function of the standard normal distribution.
Such a distribution is denoted by $SEP(\mu,\sigma,\alpha,\lambda)$.\\

The SEP reduces to the exponential power when $\lambda=0$, to the skew normal when $\alpha=2$, and to the normal when $(\lambda,\alpha)=(0,2)$. In what follows, we always consider $\alpha \le 2$, because we want our distributions to be fat-tailed. We refer the interested reader to Azzalini \cite{Azzalini86} and DiCiccio and Monti \cite{DiCiccio} for detailed results concerning this family of distributions.\\

It must be noted that $\mu$ and $\sigma$ are not the mean and the standard deviation of a $SEP(\mu,\sigma,\alpha,\lambda)$ distribution. These parameters are called ``location'' and ``scale'' instead. The even moments of a random variable $Z\sim SEP(0,1,\alpha,\lambda)$ are given by
\begin{equation}
E[Z^{2m}]= \alpha^{2m/\alpha}\Gamma((2m+1)/\alpha)/ \Gamma (1/\alpha), m\in \N,
\label{moments_SEP_even}
\end{equation}
and the odd moments by
\begin{equation}
 E[Z^{2m+1}]= \frac{2\alpha^{(2m+1)/\alpha}\lambda}{\sqrt{\pi}\Gamma(1/\alpha) (1+\lambda^2)^{s+1/2}}
\sum_{n=0}^\infty \frac{\Gamma(s+n+1/2)}{(2n+1)!!}\left( \frac{2\lambda^2}{1+\lambda^2}  \right)^n ,
\label{moments_SEP_odd}
\end{equation}
where $s=2(m+1)/\alpha$ and $(2n+1)!! := 1\cdot 3 \cdots (2n-1)\cdot (2n+1)$ -- see \cite{DiCiccio}. In particular, when $\lambda \geq 0$, $E[Z]\geq 0$.\\

Hereafter, we always assume that the cumulative distribution functions $F_0$, $F^*_0$, $F_0(\cdot;n)$, and $F_0^*(\cdot;n)$ are all of the skew exponential power type ($n = 0,\ldots,5$).\\

As far as clients are concerned, we do not have any empirical intuition
for the form of the distribution of reservation values. Therefore, by
default, we propose to use a rather naive distribution. We assume that $V_{i}$ is
Gaussian conditionally on $\Omega_i$: in the case of a ``buy'' (resp. ``sell'') order, $G_{0}(\cdot\left|\Omega_i)\right.$ (resp. $G^*_{0}(\cdot\left|\Omega_i)\right.$) is the
cumulative distribution function of a Gaussian random variable $\Nc(\nu,\tau^2)$ (resp. $\Nc(\nu^*,(\tau^*)^2)$).
When a client is interested in buying (resp. selling) a particular bond, we expect that he/she thinks that the bond is underpriced (overpriced) by the market.
In other words, we expect $\nu >0$ and $\nu^*<0$.\\

\subsection{To answer or not to answer}

For each RFQ in our database, we know how many dealers have been requested by the client. As described above, we consider two versions of our model. In the first version, we assume that all the requested dealers answer a price. In the second one, the dealers may answer or not answer (by construction, the reference dealer always answers).\\

In practice, some dealers simply do not want to answer because they are not interested in trading the requested bond, or not interested in dealing with the client who has sent the RFQ. In that case, not answering is equivalent to answering a very bad price. However, some dealers may also not answer because they are not at their desk when the RFQ pops up on their screen, or because they do not answer fast enough (the client may have accepted the price of another dealer for instance). An important issue is that we cannot distinguish between the different situations leading a dealer to not answering a RFQ. We do not observe indeed the complete sequence of quotes received by clients, and their precise timing. Therefore, we will assume that, once a RFQ is sent, the corresponding client waits for a ``reasonable'' amount of time (or until he/she has received a ``sufficient'' number of quotes) before making a decision about whether or not to transact (and with whom).\\

In our second model, we intend to measure to what extent the no-answer effect is significant. In this second model, only a subset of the requested dealers will compete for real. Therefore, we introduce a new variable corresponding to the ``effective'' number of answers. For the RFQ $i$, the ``effective'' number of answers (including the reference dealer's one) is denoted by $\tilde n_i + 1$, where
\begin{equation}
 \tilde n_i = \sum_{k=1}^{n_i} \1_{A_{k,i}=1}.
 \label{attrition_def}
 \end{equation}
Given the market and the RFQ information $\Omega_i$, we assume that the answer indicators $(A_{k,i})_k$ are mutually independent and identically distributed.
Moreover, $A_{k,i}$ is independent of $A_{k',i'}$ when $i\neq i'$, for any $k$ and $k'$ (the random variables associated with two different RFQs are independent).\\

Therefore, by~(\ref{attrition_def}), the law of $\tilde n_i$ is binomial $\mathcal{B}(n_i,p_i)$, where
$p_i := P(A_{k,i} = 1 |\Omega_i)$. In particular, $$p_{i,j} := P(\tilde n_i =j|\Omega_i) =
\binom{n_i}{j} p_i^{j} (1-p_i)^{n_i-j}, \qquad j=0,\ldots,n_i.$$

In practice, it is likely that the probability $p_i$ that a dealer answers depends on many variables. It could depend on the bond characteristics, on the time of the RFQ,\footnote{During lunch breaks, $p_i$ should be smaller.} on the market activity, on dealer or client identities, etc.\footnote{Moreover, the response process could be regarded as ``endogenous''. For instance, it could be driven by the dealers' quotes themselves, because a client can accept the price of a dealer without waiting for the other dealers to answer. We do not try to capture such effects that would make the
model significantly more complex, by inducing a feedback effect between the variable $A_{k,i}$ and the variables $W_{k',i}$, $k'\neq k$.}
In this paper, we will simply assume that $p_i$ is uniform across RFQs, or that it depends only on the number of requested dealers.\\

The model we propose can be extended to consider a more flexible law for the ``effective'' number of answers. In fact, one can even set a non-parametric distribution for $\tilde n_i$ by considering the probabilities $q_{k,i} := P(\tilde n_i = k |\Omega_i)$ for every $i$ and $k=0,\ldots,n_i$, and assuming for instance that $q_{k,i} =: q_k(n_i)$ depends on $k$ and $n_i$ only. However, since the number of unknown parameters in this extended model is a lot larger than with the above binomial specification,\footnote{One has to estimate $q_k(n_i)$, for $n_i=0,\ldots,5$ and $k=0,\ldots,n_i$.} we have only carried out estimations in the binomial case.

\section{Presentation of the dataset and estimation method}

\subsection{The dataset}

\label{Estimation}

Our work is based on an historical database of RFQs received and answered by BNP Paribas, the ``reference dealer''. If all the quotes $V_i$ and $W_{k,i}$ and the final outcomes of (a subset of) RFQs were available in this database, it would be easy to infer $F$ and $G$. Unfortunately, this is not the case. Actually, we face a partial information problem, because the amount of information we can retrieve from the RFQs is strongly constrained and limited.\\

To be specific, we get the following information with RFQ $i$:\footnote{More information is available, in particular, the identity of the client is known by the dealer before he answers a price. We have not used this information in this paper.}
\begin{itemize}
\item The outcome of the RFQ:
%\footnote{We only considered the RFQs for which the reference dealer has answered a quote.}
\begin{itemize}
\item $I_i=1$ (Done), when the RFQ resulted in a trade with the reference dealer.
\item $I_i=2$ (Traded Away), when the RFQ resulted in a trade, but with another dealer.
In that case, another discrete dummy variable $J_i\in \{1,2,3\}$ is available in the database:
\begin{itemize}
\item $J_i=1$ means ``Tied Traded Away'', \emph{i.e.}, the reference dealer has proposed exactly the same price as the winner, but has not been chosen.
\item $J_i=2$ means ``Covered'', \emph{i.e.}, the reference dealer has proposed the second best price, and was the only dealer to propose this price.
\item $J_i=3$ means ``Other Traded Away''. In that case, we have no information and this value is considered as missing. It is noteworthy that we may be covered in this case, but we do not know it.
\end{itemize}
\item $I_i=3$ (Not Traded), when the RFQ resulted in no trade.
\end{itemize}
\item The second best dealer price $C_i$, called the ``cover price'', when the reference dealer has made the deal, and when there was another answer.
\item $Y_i$, the price/quote answered by the reference dealer for this RFQ.
Note that this is the price of the deal when $I_i=1$, \emph{i.e.}, when the reference dealer has been chosen by the client.
\item $n_i$, the number of dealers requested during this RFQ, in addition to our reference dealer ($n_i\in \{ 0, \ldots, 5\}$).\footnote{We only considered the cases $n_i\ge 1$, because $n_i=0$ is specific.}
\item $Z_i$, a vector of bond characteristics.
\end{itemize}
Our dataset contains a $N$-sample of such information. After filtering, the number of ``buy'' (resp. ``sell'') RFQs is equal to $209\,069$ (resp. $272\,189$). More precisely, we have the following breakdown of RFQs in the database:\\

\begin{table}
\centering
{\footnotesize{
\begin{tabular}{|c|c|c|c|c|c|c|}
  \hline
  &$I=1$&$I=2,J=1$&$I=2,J=2$ &$I=2,J=3$&$I=3$ & \multirow{2}{*}{Total}\\
  &(Done)&(Tied Traded Away)&(Covered)&(Other Traded Away)& (Not Traded)& \\\hline
$n=1$&1845&44&664&285&2418&5256\\\hline
$n=2$&4065&157&2754&1789&4286&13051\\\hline
$n=3$&5140&259&4553&4869&5412&20233\\\hline
$n=4$&12918&975&13377&20267&13462&60999\\\hline
$n=5$&20777&1457&22706&42506&22084&109530\\\hline
Total&44745&2892&44054&69716&47662&209069\\\hline
\end{tabular}}}
\caption{Number of buy RFQs by number of other requested dealers and outcome.}
\end{table}

\begin{table}
\centering
{\footnotesize{
\begin{tabular}{|c|c|c|c|c|c|c|}
  \hline
  &$I=1$&$I=2,J=1$&$I=2,J=2$ &$I=2,J=3$&$I=3$ & \multirow{2}{*}{Total}\\
  &(Done)&(Tied Traded Away)&(Covered)&(Other Traded Away)& (Not Traded)& \\\hline
$n=1$&2125&54&915&426&2844&3520\\\hline				
$n=2$&4871&221&4047&3137&4934&12276\\\hline				
$n=3$&6226&339&6064&7954&5400&20583\\\hline				
$n=4$&14634&1212&16562&31610&12889&64018\\\hline				
$n=5$&24768&1736&28138&69020&22063&123662\\\hline				
Total&52624&3562&55726&112147&48130&272189\\\hline				
\end{tabular}}}
\caption{Number of sell RFQs by number of other requested dealers and outcome.}
\end{table}

The available information $\Omega_i$ for all market
participants includes the variables $n_i$, $Z_i$ and the standard
market information (news, etc.), in addition to CBBT prices. The information provided by a given RFQ is assumed to be independent of the information associated with other RFQs.
In particular, we assume no ``learning'' effect, \emph{i.e.}, the clients do not learn from past RFQs.
The dealers competing with the reference dealer propose prices $(W_{k,i})_k$ drawn independently from the same conditional distribution
$F(\cdot | \Omega_i)$ -- or $F^*(\cdot | \Omega_i)$ --, that depend on the bond
characteristics and on the number of dealers maybe. These
quotes and $V_{i}$ are chosen independently, given
the whole market information (no collusion).\\

Let us concatenate the information into a sample $\Sc_N =(
Y_i,n_i,C_i,I_i,J_i,Z_i)_{i=1}^N:=(\XX_i)_{i=1,\ldots,N}$.
Our goal is to estimate the
conditional distributions $F(\cdot \left|\Omega_i)\right.$, $G(\cdot \left|\Omega_i)\right.$, $F^*(\cdot \left|\Omega_i)\right.$ and $G^*(\cdot \left|\Omega_i)\right.$.\\

\subsection{Maximum likelihood inference}

Usual log-likelihood maximization methodologies can be invoked for evaluating all the unknown model parameters. We can indeed write the likelihood associated with each RFQ. Let us start with our first model where each dealer answers a price.

\subsubsection{The ``full-participation'' model}

Let us detail the likelihood associated with the sample $\Sc_N$, if we assume that each requested dealer answers a price. In what follows, we focus on the case of ``buy'' orders. Similar expressions can be obtained in the case of ``sell'' orders.\\

{\bf a. If $I_i=1$ (Done)}, then the likelihood of $\XX_i$ in the case of a ``buy'' order is
\begin{eqnarray*}
\Lc_{i|\text{buy}|\text{cover}}^{(1)}(n_i)& =& P\left(\min_{k=1,\ldots,n_i} W_{k,i} = C_i, V_{i}\geq Y_i |\Omega_i\right)\\
&=& n_i f(C_i|\Omega_i)(1-F(C_i |\Omega_i))^{n_i-1} (1-G(Y_i |\Omega_i)),
 \end{eqnarray*} if we know the cover price $C_i$ (which is larger than $Y_i$).\\

When the cover price is unknown, the likelihood of $\XX_i$ can be written as
\begin{eqnarray*}\Lc_{i|\text{buy}|\text{nocover}}^{(1)}(n_i) &=& P\left(\min_{k=1,\ldots,n_i} W_{k,i} \ge Y_i, V_{i}\geq Y_i |\Omega_i\right)\\
& =& (1-F(Y_i |\Omega_i))^{n_i} (1-G(Y_i |\Omega_i)).
\end{eqnarray*}

{\bf b. If $I_i=2$ (Traded Away)}, then the likelihood of $\XX_i$ depends on the value of $J_i$.\\

Let us first consider the dealers' quotes $(W_{k,i})_k$ corresponding to the RFQ $i$ in ascending order:\footnote{We skip here the index $i$ of the RFQ.} $W_{(1)}\leq W_{(2)} \leq \cdots \leq W_{(n_i)}$.\\

We recall that the joint probability density function of $(W_{(1)},W_{(2)})$  is
$$ f_{(1),(2)}( w_{(1)},w_{(2)} \left|\Omega_i)\right. = n_i(n_i-1)
\left(1- F(w_{(2)}\left|\Omega_i)\right.\right)^{n_i-2} f( w_{(1)} \left|\Omega_i)\right.  f( w_{(2)} \left|\Omega_i)\right.\1_{w_{(1)} \leq w_{(2)}}.$$

\begin{itemize}
\item In the ``Tied Traded Away'' case ($J_i=1$), the likelihood of $\XX_i$ in the case of a ``buy'' order is
\begin{eqnarray*}
\Lc_{i|\text{buy}}^{(2,1)}(n_i) &=& P\left(W_{(1)} = Y_i \leq V_{i}  |\Omega_i\right)\\
&=&  P\left(W_{(1)} = Y_i   |\Omega_i\right) P\left( Y_i \leq V_{i}  |\Omega_i\right) \\
&=&  n_i (1-F(Y_i \left|\Omega_i)\right.)^{n_i-1} f(Y_i |\Omega_i) (1-G (Y_i  |\Omega_i)).
\end{eqnarray*}

\item In the ``Covered'' case ($J_i=2$), the likelihood of $\XX_i$ in the case of a ``buy'' order, when $n_i\ge2$, is
\begin{eqnarray*}
\Lc_{i|\text{buy}}^{(2,2)}(n_i) &=& P\left( W_{(1)} < Y_i < W_{(2)}, W_{(1)}\leq V_{i}  |\Omega_i\right) \\
&=& E\left[ \1_{W_{(1)} < Y_i < W_{(2)}} (1-G(W_{(1)} | \Omega_i)) |\Omega_i \right] \\
&=&  \int \1_{w_{(1)} < Y_i < w_{(2)} } (1-G(w_{(1)} | \Omega_i)) f_{(1),(2)}(w_{(1)},w_{(2)} |\Omega_i) dw_{(1)}dw_{(2)}\\
&=&  n_i(n_i-1) \int_{-\infty}^{Y_i} (1-G(w_{(1)} | \Omega_i)) f( w_{(1)} |\Omega_i) dw_{(1)}\\
 &\times&\int^{+\infty}_{Y_i} \left(1- F(w_{(2)}|\Omega_i)\right)^{n_i-2}   f( w_{(2)} |\Omega_i) dw_{(2)}\\
&=&  n_i \left(1- F(Y_i|\Omega_i)\right)^{n_i-1} \int_{-\infty}^{Y_i} (1-G(w_{(1)} | \Omega_i)) f( w_{(1)} |\Omega_i) dw_{(1)}
\end{eqnarray*}
This likelihood can also be written as
$$\Lc_{i|\text{buy}}^{(2,2)}(n_i) = n_i \left(1- F(Y_i|\Omega_i)\right)^{n_i-1}$$
 $$\times\left(1- (1-F(Y_i | \Omega_i))(1-G(Y_i | \Omega_i)) - \int_{-\infty}^{Y_i} (1-F(w_{(1)} | \Omega_i)) g( w_{(1)} |\Omega_i) dw_{(1)}\right).$$
The latter formulas also apply to the case $n_i=1$:
\begin{eqnarray*}
\Lc_{i|\text{buy}}^{(2,2)}(n_i) &=& P\left( W_{(1)} < Y_i , W_{(1)}\leq V_{i}  |\Omega_i\right) \\
&=&  \int \1_{w_{(1)} < Y_i} (1-G(w_{(1)} | \Omega_i)) f(w_{(1)} |\Omega_i) dw_{(1)}\\
&=&\! 1\!-\! (1-F(Y_i | \Omega_i))(1-G(Y_i | \Omega_i))\! -\! \int_{-\infty}^{Y_i} (1-F(w_{(1)} | \Omega_i)) g( w_{(1)} |\Omega_i) dw_{(1)}.
\end{eqnarray*}

\item In the ``Other Traded Away'' case ($J_i=3$), the likelihood of $\XX_i$ in the case of a ``buy'' order is
\begin{eqnarray*}
\Lc_{i|\text{buy}}^{(2,3)}(n_i)&=& P\left(\min_{k=1,\ldots,n_i} W_{k,i} \leq \min( V_{i}, Y_i) |\Omega_i\right) \\
&=& E\left[\left(  1 - E\left[\1_{\min_{k=1,\ldots,n_i} W_{k,i} \geq \min (V_{i},Y_i) } | V_{i}, \Omega_i\right]  \right) |\Omega_i \right] \\
&=& E\left[ \left(  1 - (1-F( \min(V_{i} , Y_i) | \Omega_i ))^{n_i}   \right) |\Omega_i \right] \\
\end{eqnarray*}
\begin{eqnarray*}
&=& \int  \left(  1 - (1-F( \min(v,Y_i) | \Omega_i ))^{n_i}   \right) g(v |\Omega_i)dv\\
&=& \int_{-\infty}^{Y_i}  \left(  1 - (1-F(v | \Omega_i ))^{n_i}   \right) g(v |\Omega_i)dv\\
 && + \left(  1 - (1-F(Y_i | \Omega_i ))^{n_i}   \right) \left( 1 -  G(Y_i |\Omega_i)\right)\\
&=& 1 - (1-F(Y_i | \Omega_i ))^{n_i} \left( 1 -  G(Y_i |\Omega_i)\right)\\
  &-& \int_{-\infty}^{Y_i}  (1-F(v | \Omega_i ))^{n_i} g(v |\Omega_i)dv\\
\end{eqnarray*}
\end{itemize}

{\bf c. If $I_i=3$ (Not Traded)}, then the likelihood of $\XX_i$ in the case of a ``buy'' order is
$$ \Lc_{i|\text{buy}}^{(3)}(n_i) =  P\left(\min_{k=1,\ldots,n_i} W_{k,i} \geq V_{i}, Y_i \geq V_{i}
|\Omega_i\right) = \int_{-\infty}^{Y_i}  (1-F(v |\Omega_i))^{n_i} g(v |\Omega_i) dv.$$

It is noteworthy that these likelihoods only involve integrals of the form
$$\int_{-\infty}^{Y_i}  (1-F(v |\Omega_i))^{k} g(v |\Omega_i) dv, \qquad k \in \lbrace 1, \ldots, 5\rbrace.$$
From a numerical standpoint, we will only need to approximate these integrals (and the integral which defines $F$).

\subsubsection{The ``partial-participation'' model}

We now consider the second version of our model in which the requested dealers may answer or not answer. With the notations introduced in Section 2.3, we have (we focus on ``buy'' orders, but ``sell'' orders can be tackled similarly):\\

{\bf a. If $I_i=1$ (Done)}, then the likelihood of $\XX_i$ in the case of a ``buy'' order is
$$ \Lc_{i|\text{buy}|\text{cover}}^{(1)} = \sum_{j=1}^{n_i} p_{i,j}  \Lc_{i|\text{buy}|\text{cover}}^{(1)}(j),$$ if we know the cover price.\\

When the cover price is unknown, it is either because the reference dealer was the only one to answer, or because the cover price has not been recorded. The likelihood of $\XX_i$ can be written as
$$ \Lc_{i|\text{buy}|\text{nocover}}^{(1)} = \sum_{j=0}^{n_i} p_{i,j} \Lc_{i|\text{buy}|\text{nocover}}^{(1)}(j) .$$

{\bf b. If $I_i=2$ (Traded Away)}, then the likelihood of $\XX_i$ depends on the value of $J_i$.\\

\begin{itemize}
\item In the ``Tied Traded Away'' case ($J_i=1$), the likelihood of $\XX_i$ in the case of a ``buy'' order is
\begin{eqnarray*}
\Lc_{i|\text{buy}}^{(2,1)}&=& \sum_{j=1}^{n_i}  p_{i,j}  \Lc_{i|\text{buy}}^{(2,1)}(j).
\end{eqnarray*}

\item In the ``Covered'' case ($J_i=2$), the likelihood of $\XX_i$ in the case of a ``buy'' order is
\begin{eqnarray*}
\Lc_{i|\text{buy}}^{(2,2)}&=& \sum_{j=1}^{n_i} p_{i,j} \Lc_{i|\text{buy}}^{(2,2)}(j).
\end{eqnarray*}

\item In the ``Other Traded Away'' case ($J_i=3$), the likelihood of $\XX_i$ in the case of a ``buy'' order is
\begin{eqnarray*}
\Lc_{i|\text{buy}}^{(2,3)}&=& \sum_{j=1}^{n_i} p_{i,j}  \Lc_{i|\text{buy}}^{(2,3)}(j)
\end{eqnarray*}
\end{itemize}

{\bf c. If $I_i=3$ (Not Traded)}, then the likelihood of $\XX_i$ in the case of a ``buy'' order is
$$ \Lc_{i|\text{buy}}^{(3)}= \sum_{j=0}^{n_i} p_{i,j} \Lc_{i|\text{buy}}^{(3)}(j)$$

\section{Empirical results}
\label{empirics}

In order to estimate the value of the parameters in the full-participation model and in the partial-participation model, we maximize the log-likelihood associated with our sample and under our model specification. For maximizing this log-likelihood, we need first to be able to compute the log-likelihood associated with each RFQ (each line of the database) for given values of the parameters $(\mu,\sigma,\alpha,\lambda,\nu,\tau)$ and $(\mu^*,\sigma^*,\alpha^*,\lambda^*,\nu^*,\tau^*)$  -- and given values of $(p_i)_i$ in the partial-participation model. As we have seen above, the difficulty lies in (i) computing the cumulative distribution function $F$ defined by
$$F: y \mapsto \int_{-\infty}^y f(v) dv,$$ and (ii) computing functions of the form
$$y \mapsto \int_{-\infty}^{y}  (1-F(v))^{k} g(v) dv, \qquad k \in \lbrace 1, \ldots, 5\rbrace.$$
For numerically approximating these functions, we first rescaled the integrands to the definition interval $(-1,1)$ (by means of $\tanh$ transforms) and approximated them with  Chebyshev polynomials of the first kind. Then we used classical results on Chebyshev polynomials for finding the antiderivatives -- see \cite{nr} for a complete description of this methodology. For maximizing the log-likelihood of the sample, we used Powell's method which does not require to formally differentiate the log-likelihood with respect to the parameters -- see \cite{p,nr}.

\subsection{The full-participation model}

\subsubsection{General results}

We present the estimation of the parameters in the full-participation model for ``buy'' and ``sell'' orders separately. The estimated values of the parameters are exhibited in Table \ref{res_buy_sell}.\\

\begin{figure}[H]
  \centering
  \includegraphics[width=270pt]{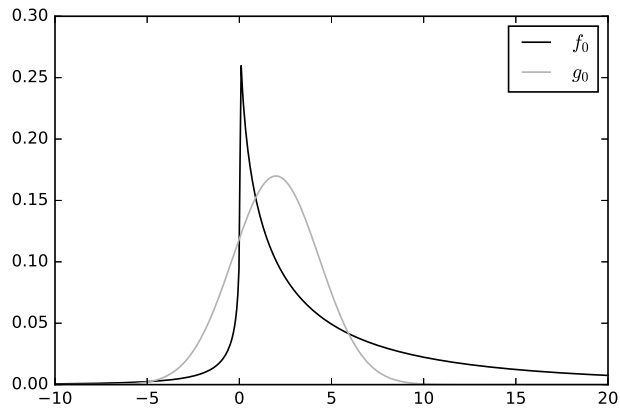}\\
  \caption{Density functions $f_0$ and $g_0$ associated with $F_0$ and $G_0$ respectively, in the case of ``buy'' orders. Black line: SEP distribution for the dealers' quotes. Gray line: Gaussian distribution for the clients' reservation prices.}\label{allbuy}
\end{figure}

\begin{figure}[H]
  \centering
  \includegraphics[width=270pt]{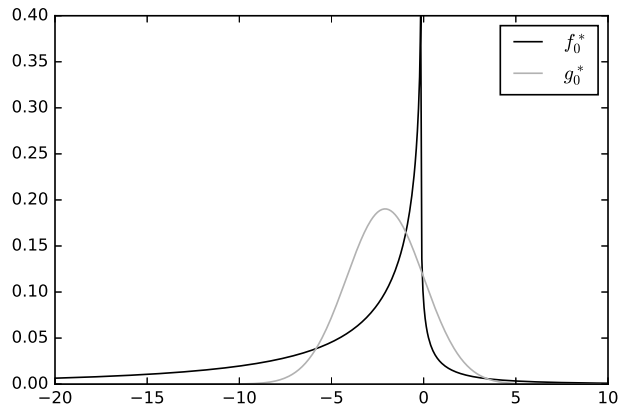}\\
  \caption{Density functions $f^*_0$ and $g^*_0$ associated with $F^*_0$ and $G^*_0$ respectively, in the case of ``sell'' orders. Black line: SEP distribution for the dealers' quotes. Gray line: Gaussian distribution for the clients' reservation prices.}\label{allsell}
\end{figure}

\begin{table}[H]
\begin{center}
{\footnotesize
\begin{tabular}{|c|c|c|c|c|c|c|} \hline
&\multicolumn{4}{|c|}{\text{Dealers}} & \multicolumn{2}{|c|}{\text{Clients}}  \\ \hline
\multirow{2}{*}{``Buy'' orders} &$\alpha$ & $\lambda$ & $\mu$ & $\sigma$ & $\nu$ & $\tau$   \\ \cline{2-7}
& $0.445$&  $0.787$&  $0.0411$&  $5.01$&  $1.98$&   $2.35$ \\ \hline
\multirow{2}{*}{``Sell'' orders} &$\alpha^*$ & $\lambda^*$ & $\mu^*$ & $\sigma^*$ & $\nu^*$ & $\tau^*$   \\ \cline{2-7}
& $0.388$&  $-0.612$&  $-0.149$&  $4.09$&  $-2.09$&   $2.10$ \\ \hline
\end{tabular} \caption{Estimation of the full-participation model parameters for all ``buy'' and ``sell'' orders.}\label{res_buy_sell} }
\end{center}
\end{table}
\vspace{-7mm}
The corresponding probability density functions for dealers' quotes and clients' reservation prices are plotted in Figures~\ref{allbuy} and~\ref{allsell}. We see that the distributions of the dealers' quotes are clearly asymmetric. To understand the rationale of this empirical result, let us consider the case of a ``buy'' RFQ. For the dealers, there is almost no difference between answering a high price and a very high price: in both cases, the price will be too high to be accepted by the client, and there will be no trade. However, there is a significant difference between answering a low price and a very low price: in both cases a trade may occur, and the trade price is the price answered by the dealer. The same reasoning applies in the case of a ``sell'' RFQ, \emph{mutatis mutandis}. This explains the skewness ($\lambda^* < 0 < \lambda$) of the distributions of the dealers' quotes.\\

These distributions are also heavy-tailed (see the value of $\alpha$ and $\alpha^*$ in Table 1). This is an important feature, but it has to be considered with care. A reason why there is a fat tail on the right-hand side (resp. left-hand side) for ``buy'' (resp. ``sell'') RFQs is indeed linked to the assumption in our first model: every requested dealer answers a price. In practice however, some of the requested dealers do not actually answer, and this is somehow equivalent to answering very conservative prices, hence an effect on the right-hand side (resp. left-hand side) tail.\\
\vspace{-4mm}
\vspace{-5mm}

\begin{figure}[H]
  \centering
  \includegraphics[width=260pt]{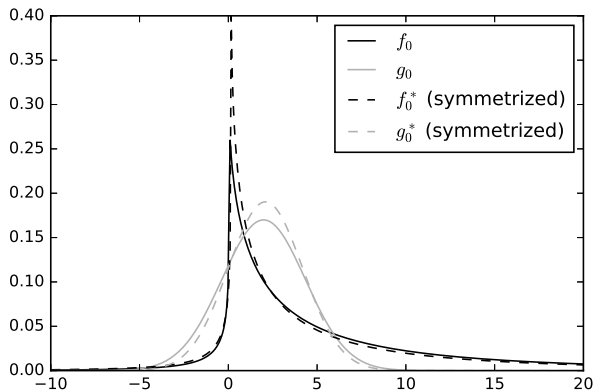}\\
  \caption{Comparison of the distributions of dealers' quotes and clients' reservation prices for ``buy'' and ``sell'' RFQs. Black: SEP distributions for the dealers. Gray: Gaussian distributions for the clients.
  Solid lines represent the case of ``buy'' RFQs. Dashed lines represent the case of ``sell'' RFQs, after symmetrization.}\label{allsellreversedbuy}
\end{figure}

\vspace{-4mm}
A visual comparison between the buy and sell cases is made possible by changing the sign, in the ``sell'' case, of the location and asymmetry parameters $\mu^*$
and $\lambda^*$ of the SEP distribution of dealers' quotes, and the mean $\nu^*$ of the Gaussian distribution of clients' reservation values -- see Figure~\ref{allsellreversedbuy}.\\
\vspace{-5mm}

\vspace{-3mm}

It is noteworthy that the probability density function for the dealers' quotes is more ``spiky'' in the ``sell'' case than in the ``buy'' one, but there is overall not much difference between ``buy'' and ``sell'' orders for both dealers and clients. In particular $\nu \simeq -\nu^* \simeq 2 > 0$, and this means that on average, a client sending a ``buy'' (resp. ``sell'') order thinks that the CBBT mid-price underestimates (resp. overestimates) the fair value of the bond by an amount equal to the CBBT bid-ask spread -- with a standard deviation $\tau \simeq \tau^* \simeq 2$.

\subsubsection{The influence of competition}

We now turn to the results obtained with our log-likelihood maximization procedure on the different subsamples of RFQs corresponding to different numbers of requested dealers. In other words, instead of estimating $F_0$, $F^*_0$, $G_0$ and $G_0^*$, we estimate the distributions $F_0(\cdot;n)$, $F_0^*(\cdot;n)$, $G_0(\cdot;n)$, and $G_0^*(\cdot;n)$, defined in Equations (\ref{Fn}) and (\ref{Gn}) -- $n \in \{1, \ldots, 5\}$.\\

It is likely that the behaviors of dealers and clients depend on the level of competition. In what follows, we aim at answering questions such as: (i) how does the distribution of answered quotes depend on the number of dealers requested?, (ii) are the distributions of clients' reservation prices similar for clients requesting a few dealers and clients requesting a lot of dealers?, etc.\\

We exhibit in Tables~\ref{res_buy_all_ni} and \ref{res_sell_all_ni}  the estimated values of the parameters characterizing the SEP distributions of the dealers' quotes, and the Gaussian distributions of the clients' reservation prices, for ``buy'' and ``sell'' RFQs with different numbers of requested dealers.\\

\vspace{-4mm}

\begin{table}[H]
\begin{center}
{\footnotesize
\begin{tabular}{|l|c|c|c|c|c|c|c|} \hline
  & \multicolumn{4}{|c|}{\text{Dealers}} & \multicolumn{2}{|c|}{\text{Clients}}  \\ \hline
& $\alpha$ & $\lambda$ & $\mu$ & $\sigma$ & $\nu$ & $\tau$   \\ \hline
$n_i=1$ & 0.811&  0.694 &  0.120&  1.41&  0.896&  4.56\\
$n_i=2$ & 0.555&  0.721&   0.130&  2.58&   1.73&  2.97 \\
$n_i=3$ &  0.444&    0.670&  0.166&  3.81&  2.06&  2.84\\
$n_i=4$ &  0.377&  0.651&  0.120&  4.90&  3.25&  4.16\\
$n_i=5$ &  0.450&  0.876& -0.0432&  5.72&   2.08&  2.27\\\hline
\end{tabular} \caption{Estimation of the model parameters for ``buy'' orders. The subsamples correspond to RFQs with different (fixed) numbers of dealers ($n_i=1$ to $5$, \emph{i.e.}, from
two to six dealers).}\label{res_buy_all_ni} }
\end{center}
\end{table}

\vspace{-6mm}

\begin{table}[H]
\begin{center}
{\footnotesize
\begin{tabular}{|l|c|c|c|c|c|c|c|} \hline
  & \multicolumn{4}{|c|}{\text{Dealers}} & \multicolumn{2}{|c|}{\text{Clients}}  \\ \hline
  & $\alpha^*$ & $\lambda^*$ & $\mu^*$ & $\sigma^*$ & $\nu^*$ & $\tau^*$   \\ \hline
$n_i=1$ & 0.597&  -0.330&  -0.286&  1.16&  -1.24& 7.41\\
$n_i=2$ &  0.490&  -0.576&  -0.137&  2.09&  -2.07& 3.19\\
$n_i=3$ &  0.420&  -0.574&  -0.166&  3.08&  -2.40& 2.73\\
$n_i=4$ &  0.348&  -0.545& -0.169&  4.12&  -2.48&   2.50\\
$n_i=5$ &  0.382&  -0.660&  -0.111&  4.67&  -2.13&    1.94\\
\hline
\end{tabular} \caption{Estimation of the model parameters for ``sell'' orders. The subsamples correspond to RFQs with different (fixed) numbers of dealers ($n_i=1$ to $5$, \emph{i.e.}, from
two to six dealers).}\label{res_sell_all_ni} }
\end{center}
\end{table}

\vspace{-6mm}

\begin{figure}[H]
  \centering
  \includegraphics[width=270pt]{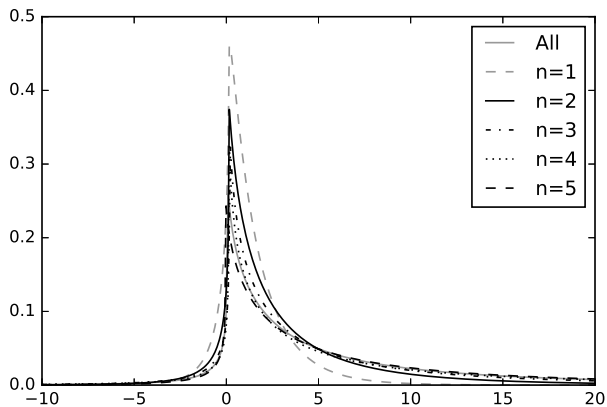}\\
  \caption{SEP distributions $(f_0(\cdot;n))_n$ for the dealers' quotes on the subsamples of ``buy'' RFQs. Gray dashed line: $n=1$. Black solid line: $n=2$. Black dash-dotted line: $n=3$. Black dotted line: $n=4$.
  Black dashed line: $n=5$.  Gray solid line: all ``buy'' RFQs.}\label{sep23456buy}
\end{figure}

\begin{figure}[H]
  \centering
  \includegraphics[width=270pt]{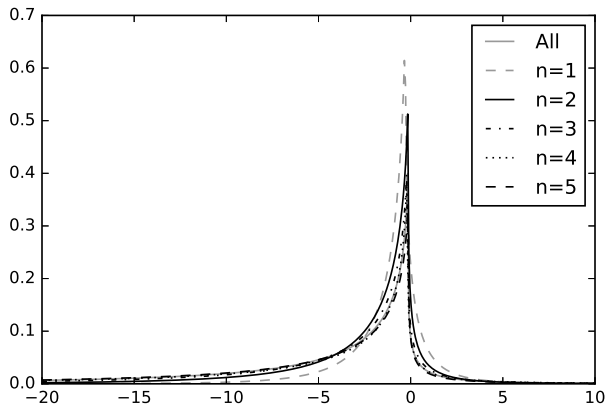}\\
  \caption{SEP distributions $(f^*_0(\cdot;n))_n$ for the dealers' quotes on the subsamples of ``sell'' RFQs. Gray dashed line: $n=1$. Black solid line: $n=2$. Black dash-dotted line: $n=3$. Black dotted line: $n=4$.
  Black dashed line: $n=5$.  Gray solid line:  all ``sell'' RFQs.}\label{sep23456sell}
\end{figure}

The distributions of dealers' quotes for the different values of the number of requested dealers are plotted in Figures~\ref{sep23456buy} and \ref{sep23456sell}.

We clearly see that the distribution of dealers' quotes depends on the number of requested dealers in a very specific and ordered way. In particular, in the case of ``buy'' (resp. ``sell'') RFQs, we see that the heaviness of the right-hand side (resp. left-hand side) tail increases with the number of dealers in competition. Graphically, we also see that the smaller the number of competitors, the higher the probability a dealer chooses a price below (resp. above) the CBBT mid-price when he/she answers a ``buy'' (resp. ``sell'') order. In other words, in the first model, it seems that the larger the number of dealers in competition, the more conservative their answered quotes.\\

The latter monotonicity property is surprising and could be explained by the behavior of dealers. When a dealer receives a RFQ sent to a few dealers only, he/she may think that his/her effort to propose a good price will lead to a deal, because competition is not strong. Conversely, in the case of a RFQ sent to many dealers, he/she may think that there is little chance for him/her to be chosen, and therefore no reason to spend time choosing a relevant non-conservative price. This can be regarded as some form of ``discouragement'' effect. Another important -- and in fact better (see the results of the ``partial-participation'' model) -- explanation is related to dealers who do not answer: the larger the number of requested dealers, the higher the probability that a dealer does not have time to answer, for instance because the client has traded with an early-answerer. In other words, the larger the number of requested dealers, the more important this ``no-answer'' effect. In our first (``full-participation'') model, this effect artificially leads to an increase in the estimated probability of conservative answered quotes -- whence our extended (partial-participation) model.\\

\begin{figure}[H]
  \centering
  \includegraphics[width=270pt]{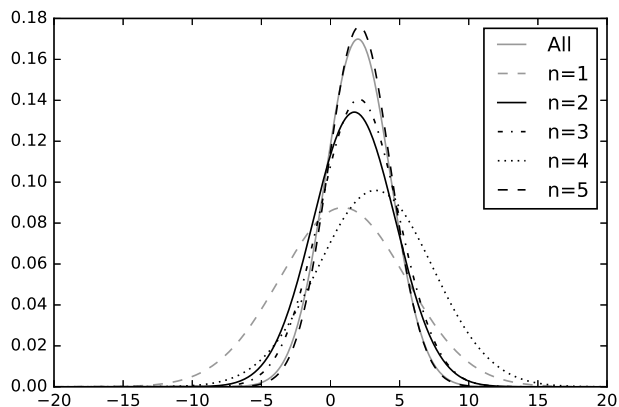}\\
  \caption{Gaussian distributions $(g_0(\cdot;n))_n$ for the clients' reservation values on the subsamples of ``buy'' RFQs.
  Gray dashed line: $n=1$. Black solid line: $n=2$. Black dash-dotted line: $n=3$. Black dotted line: $n=4$.
  Black dashed line: $n=5$.  Gray solid line:  all ``buy'' RFQs.}\label{clientgauss23456buy}
\end{figure}

\begin{figure}[t]
  \centering
  \includegraphics[width=270pt]{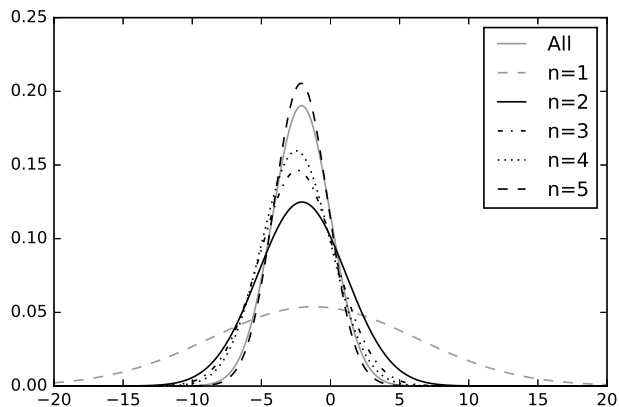}\\
  \caption{Gaussian distributions $(g^*_0(\cdot;n))_n$ for the clients' reservation values on the subsamples of ``sell'' RFQs.
  Gray dashed line: $n=1$. Black solid line: $n=2$. Black dash-dotted line: $n=3$. Black dotted line: $n=4$.
  Black dashed line: $n=5$.  Gray solid line:  all ``sell'' RFQs.}\label{clientgauss23456sell}
\end{figure}

What we see in Tables \ref{res_buy_all_ni} and \ref{res_sell_all_ni}, and in Figures~\ref{clientgauss23456buy} and \ref{clientgauss23456sell}, is that, as far as clients are concerned, the main influence of competition is on the standard deviation (or equivalently the variance) of their reservation price (the case $n_i=4$ for ``buy'' orders seems specific). There seems to be more homogeneity amongst clients who have requested a lot of dealers than amongst clients who have only requested few dealers. \\

In order to understand this effect, we need to understand why clients do not always request 6 dealers. First, there may be less than 6 dealers streaming prices on a specific bond. In that case, the bond is certainly illiquid and it is not surprising to observe a wider range of reservation values (even normalized by the CBBT bid-ask spread) for that bond. Second, a client willing to buy or sell a large notional of bonds may only request a few dealers to avoid information leakage. Then, the reservation value for the bond may be far from the CBBT mid-price for that notional, hence the large variance. In addition to the monotonicity effect, the $n_i=1$ case seems to be specific. In addition to the above explanations, it is important to know that some (informal) agreements between clients and dealers are reached outside of the MD2C platform: an interesting price may be proposed by a dealer to a client by telephone or chat, and then the formal agreement reached on the platform after the client has requested two dealers for proving he/she received best execution.\\

\subsection{The partial-participation model}
\label{partial_part_model}

We now turn to the estimation of the parameters in the partial-participation model for ``buy'' and ``sell'' orders separately. The estimated values of the parameters are exhibited in Table \ref{res_buy_sellbin}.\\

\begin{table}[H]
\begin{center}
{\footnotesize
\begin{tabular}{|c|c|c|c|c|c|c|c|c|c|} \hline
&\multicolumn{7}{|c|}{\text{Dealers}} & \multicolumn{2}{|c|}{\text{Clients}}  \\ \hline
\multirow{2}{*}{``Buy'' orders} &$\alpha$ & $\lambda$ & $\mu$ & $\sigma$ & $\text{mean}$& \text{std. dev.}  & $p$ & $\nu$ & $\tau$   \\ \cline{2-10}
& $0.735$&  $0.179$&  $0.424$&  $0.906$& $0.748$ & $1.60$ & $0.400$ &  $1.72$& $1.92$\\ \hline
\multirow{2}{*}{``Sell'' orders} &$\alpha^*$ & $\lambda^*$ & $\mu^*$ & $\sigma^*$ & $\text{mean}$& \text{std. dev.} & $p^*$ & $\nu^*$ & $\tau^*$   \\ \cline{2-10}
& $0.665$& $-0.103$& $-0.418$&  $0.794$& $-0.605$ & $1.56$ & $0.424$ &$-1.80$& $1.68$ \\ \hline
\end{tabular} \caption{Estimation of the partial-participation model parameters for all ``buy'' and ``sell'' orders.}\label{res_buy_sellbin} }
\end{center}
\end{table}

In the partial-participation model, each requested dealer answers with probability $p$ (or~$p^*$). We see in Table~\ref{res_buy_sellbin} that the estimated values of the probabilities $p$ and $p^*$ are around $0.4$, and therefore that there is $40\%$ chance that a requested dealer actually answers -- far from the implicit hypothesis $p=p^*=1$ of the full-participation model.\\

\begin{figure}[H]
  \centering
  \includegraphics[width=270pt]{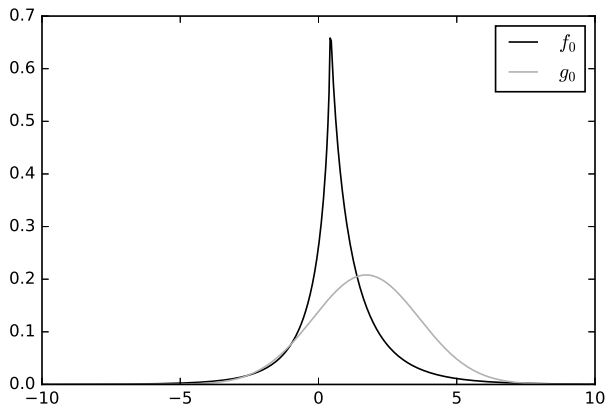}\\
  \caption{Density functions $f_0$ and $g_0$ associated with $F_0$ and $G_0$ respectively, in the case of ``buy'' orders. Black line: SEP distribution for the dealers' quotes. Gray line: Gaussian distribution for the clients' reservation prices (partial-participation model).}\label{allbuybin}
\end{figure}

\begin{figure}[H]
  \centering
  \includegraphics[width=270pt]{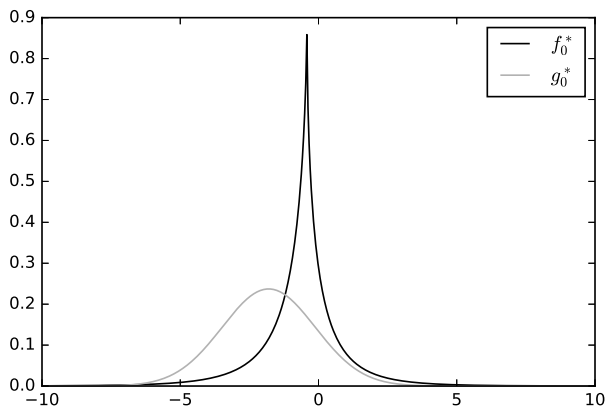}\\
  \caption{Density functions $f^*_0$ and $g^*_0$ associated with $F^*_0$ and $G^*_0$ respectively, in the case of ``sell'' orders. Black line: SEP distribution for the dealers' quotes. Gray line: Gaussian distribution for the clients' reservation prices (partial-participation model).}\label{allsellbin}
\end{figure}

As far as dealers' quotes are concerned, the SEP distributions are less heavy-tailed, as we have $\alpha \simeq \alpha^* \simeq 0.7$ instead of values around $0.4$ in the full-participation model (see Table~\ref{res_buy_sell}). It is in fact clear, in Figures~\ref{allbuybin} and~\ref{allsellbin}, that the partial-participation model captures a very important effect: the heaviness of the right-hand side tail in Figure \ref{allbuy} and of the left-hand side tail in Figure \ref{allsell} were mainly due to the absence of answer from some dealers. Nevertheless, we see that the asymmetry property $\lambda^* < 0 < \lambda$ is still valid in this second model, and certainly related to some of the effects discussed in Section 4.1.1. In particular, there is still more economic difference between an aggressive price and a very aggressive price than between a conservative price and a very conservative price.\\

We also see in Table \ref{res_buy_sellbin} that the mean value of dealers' quotes is positive (resp. negative) in the case of a ``buy'' (resp. ``sell'') order. This means that dealers do not propose, on average, prices that are too aggressive. More precisely, we have that the mean value of dealers' quotes is around 30-40\% of the CBBT bid-ask spread above (resp. below) the CBBT mid-price in the case of ``buy'' (resp. ``sell'') orders. As far as clients are concerned, $\nu$ and $-\nu^*$ are almost equal and  $10\%$ below the values estimated in the first model. This means that, on average, a client sending a ``buy'' (resp. ``sell'') order thinks that the CBBT mid-price underestimates (resp. overestimates) the fair value of the bond by an amount equal to $90\%$ of the CBBT bid-ask spread. Overall, on average, clients are more convinced of potential underpricings/overpricings than dealers, the latter relying more on CBBT prices (by default).\\

\begin{figure}[H]
  \centering
  \includegraphics[width=290pt]{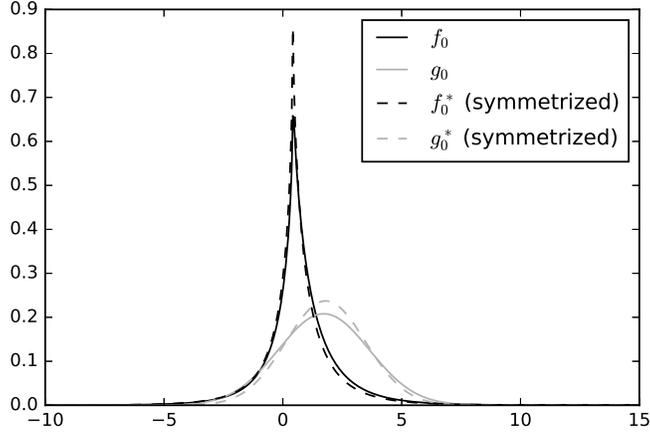}\\
  \caption{Comparison of the distributions of dealers' quotes and clients' reservation prices for ``buy'' and ``sell'' RFQs. Black: SEP distributions for the dealers. Gray: Gaussian distributions for the clients.
  Solid lines represent the case of ``buy'' RFQs. Dashed lines represent the case of ``sell'' RFQs, after symmetrization (partial-participation model).}\label{allsellreversedbuybin}
\end{figure}

As in Section 4.1.1, a visual comparison between the buy and sell cases is given by changing the sign, in the ``sell'' case, of the location and asymmetry parameters $\mu^*$
and $\lambda^*$ of the SEP distribution of dealers' quotes and the mean $\nu^*$ of the Gaussian distribution of clients' reservation values -- see Figure~\ref{allsellreversedbuybin}.\\

Once again, the probability density function for the dealers' quotes is relatively more ``spiky'' in the case of ``sell'' RFQs than in the case of ``buy'' RFQs, but there is overall not much difference between ``buy'' and ``sell'' orders.

\subsubsection{The influence of competition}
\label{PPModel_compet}
We now turn to the results obtained with our log-likelihood maximization procedure on the different subsamples of RFQs corresponding to different numbers of requested dealers.\\

We exhibit in Tables~\ref{res_buy_all_nibin} and \ref{res_sell_all_nibin}  the estimated values of the parameters characterizing the SEP distributions of the dealers' quotes, and the Gaussian distributions of the clients' reservation prices, for ``buy'' and ``sell'' RFQs with different numbers of requested dealers.

\begin{table}[H]
\begin{center}
{\footnotesize
\begin{tabular}{|l|c|c|c|c|c|c|c|c|c|c|} \hline
  & \multicolumn{7}{|c|}{\text{Dealers}} & \multicolumn{2}{|c|}{\text{Clients}}  \\ \hline
& $\alpha$ & $\lambda$ & $\mu$ & $\sigma$ & $\text{mean}$& \text{std. dev.} & $p(n_i)$ & $\nu$ & $\tau$   \\ \hline
$n_i=1$ & 0.778&  0.597&  0.208&  1.33& 1.36 & 1.96 & 1.00 & 0.890& 4.60  \\
$n_i=2$ & 0.795&  0.470&  0.285 &  1.17& 1.14 & 1.77 & 0.655 & 1.65& 2.71  \\
$n_i=3$ & 0.722&  0.247 &  0.440&  1.01& 0.933 & 1.78 &0.498 & 1.79& 2.28\\
$n_i=4$ & 0.713&  0.129&  0.480&  0.888& 0.720 & 1.62 &0.417 & 2.19& 2.53\\
$n_i=5$ & 0.738&  0.141 &  0.409&  0.840& 0.647 & 1.49 &0.351 & 1.64& 1.65\\ \hline
\end{tabular} \caption{Estimation of the model parameters for ``buy'' orders. The subsamples correspond to RFQs with different (fixed) numbers of dealers ($n_i=1$ to $5$, \emph{i.e.}, from
two to six dealers).}\label{res_buy_all_nibin} }
\end{center}
\end{table}

\vspace{-6mm}

\begin{table}[H]
\begin{center}
{\footnotesize
\begin{tabular}{|l|c|c|c|c|c|c|c|c|c|c|} \hline
  & \multicolumn{7}{|c|}{\text{Dealers}} & \multicolumn{2}{|c|}{\text{Clients}}  \\ \hline
  & $\alpha^*$ & $\lambda^*$ & $\mu^*$ & $\sigma^*$ & $\text{mean}$& \text{std. dev.} & $p^*(n_i)$ & $\nu^*$ & $\tau^*$   \\ \hline
$n_i=1$ &  0.599& -0.350& -0.264& 1.16& -1.18 & 2.39 & 0.998 &-1.24& 7.35 \\
$n_i=2$ &  0.660& -0.368& -0.249& 1.03& -1.01 & 1.90&  0.695 &-1.91& 2.81 \\
$n_i=3$ &  0.647& -0.132& -0.445& 0.865& -0.712 & 1.74&  0.522 &-1.98& 2.09 \\
$n_i=4$ &  0.671& -0.0881& -0.418&  0.796& -0.577 & 1.55&  0.437 &-2.04& 1.89 \\
$n_i=5$ &   0.681&  -0.0676& -0.409& 0.719& -0.518& 1.38& 0.373 &-1.72& 1.44 \\
\hline
\end{tabular} \caption{Estimation of the model parameters for ``sell'' orders. The subsamples correspond to RFQs with different (fixed) numbers of dealers ($n_i=1$ to $5$, \emph{i.e.}, from
two to six dealers).}\label{res_sell_all_nibin} }
\end{center}
\end{table}

An important result is that the probability that a requested dealer answers a RFQ depends on the number of dealers requested. In the case of two requested dealers, we estimate that the probability of both dealers answering the request is very close to $1$. In fact, what probably happens when a client requests two dealers is that he/she waits for the two answers before making a decision (for instance for proving he/she received best execution). Then, the larger the number of requested dealers, the smaller the probability that each dealer answers.\\

This effect can be explained by at least two distinct phenomenons. First, there may be an auto-censorship effect on the dealers' side when there is a lot of competition. Dealers in competition with other dealers may feel discouraged and simply do not answer (a similar effect was discussed in Section 4.1.2). Another, and certainly more relevant, phenomenon is related to clients' behavior. A client may choose not to wait for the answers of all requested dealers and make a decision beforehand, in particular in the Done and Traded Away cases. An interesting point is that the theoretical\footnote{We assume here that BNPP has the same probability to answer as the others.} average number of answers $(n_i+1)\cdot p(n_i)$ (or equivalently $(n_i+1) \cdot p^*(n_i)$) is always around $2$ (in fact slightly above $2$ for large values of $n_i$). This means that clients often wait for only two answers before dealing.\\

\begin{figure}[H]
  \centering
  \includegraphics[width=290pt]{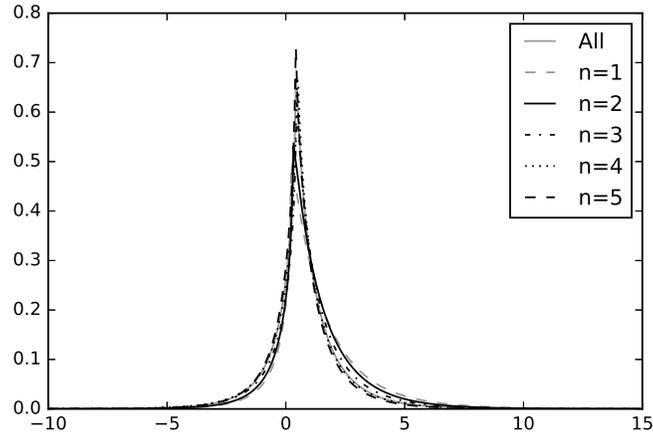}\\
  \caption{SEP distributions $(f_0(\cdot;n))_n$ for the dealers' quotes on the subsamples of ``buy'' RFQs. Gray dashed line: $n=1$. Black solid line: $n=2$. Black dash-dotted line: $n=3$. Black dotted line: $n=4$.
  Black dashed line: $n=5$.  Gray solid line: all ``buy'' RFQs (partial-participation model).}\label{sep23456buybin}
\end{figure}

\begin{figure}[H]
  \centering
  \includegraphics[width=290pt]{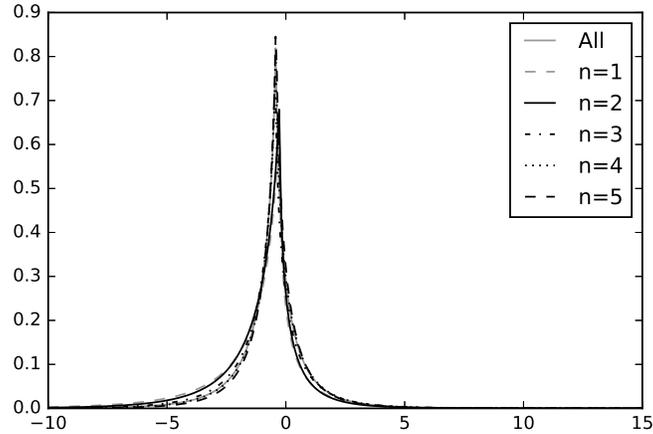}\\
  \caption{SEP distributions $(f^*_0(\cdot;n))_n$ for the dealers' quotes on the subsamples of ``sell'' RFQs. Gray dashed line: $n=1$. Black solid line: $n=2$. Black dash-dotted line: $n=3$. Black dotted line: $n=4$.
  Black dashed line: $n=5$.  Gray solid line:  all ``sell'' RFQs (partial-participation model).}\label{sep23456sellbin}
\end{figure}

The distributions of dealers' quotes for the different values of the number of requested dealers are plotted in Figures~\ref{sep23456buybin} and \ref{sep23456sellbin}.\\

We clearly see that the monotonicity property of the full-participation model has been captured by the no-answer effect. Graphically, it appears indeed that the quotes answered by the dealers (who answer) do not depend strongly on the number of requested dealers. In fact, we see in Tables 5 and 6 that the monotonicity is reversed: the mean of dealers' quotes distribution decreases (resp. increases) when $n_i$ increases for ``buy'' (resp. ``sell'') RFQs. This indicates that there is no discouragement effect among the dealers who have decided to compete effectively, but rather the opposite. To be specific, discouragement may only exists through dealer auto-selection, but not through their quote distributions. Therefore, globally, the larger the number of requested dealers, the stronger the competition between the competing dealers.\\

\begin{figure}[H]
  \centering
  \includegraphics[width=290pt]{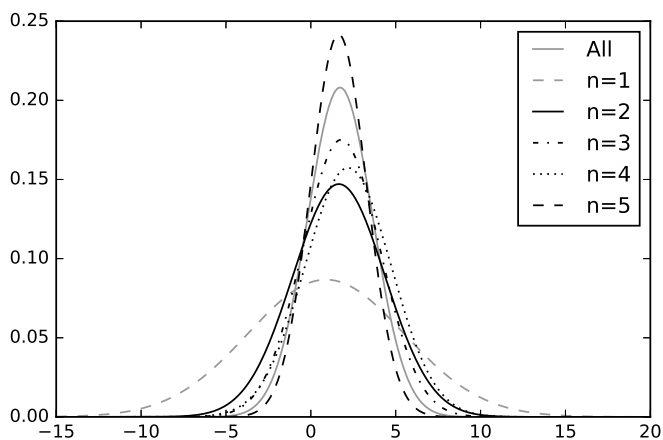}\\
  \caption{Gaussian distributions $(g_0(\cdot;n))_n$ for the clients' reservation values on the subsamples of ``buy'' RFQs.
  Gray dashed line: $n=1$. Black solid line: $n=2$. Black dash-dotted line: $n=3$. Black dotted line: $n=4$.
  Black dashed line: $n=5$.  Gray solid line:  all ``buy'' RFQs.}\label{clientgauss23456buybin}
\end{figure}

\begin{figure}[H]
  \centering
  \includegraphics[width=290pt]{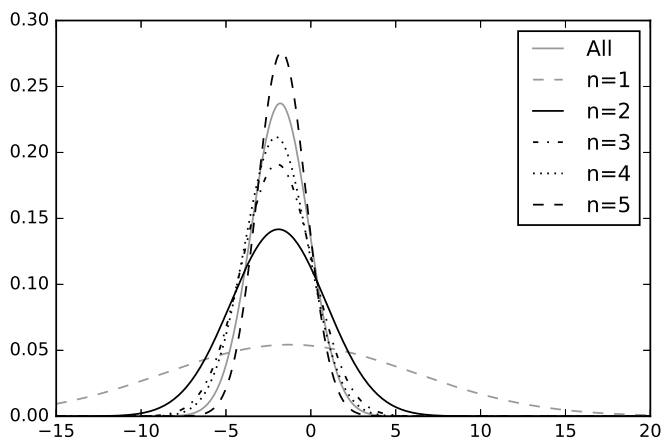}\\
  \caption{Gaussian distributions $(g^*_0(\cdot;n))_n$ for the clients' reservation values on the subsamples of ``sell'' RFQs.
  Gray dashed line: $n=1$. Black solid line: $n=2$. Black dash-dotted line: $n=3$. Black dotted line: $n=4$.
  Black dashed line: $n=5$.  Gray solid line:  all ``sell'' RFQs.}\label{clientgauss23456sellbin}
\end{figure}

The distributions of clients' reservation prices for the different values of the number of requested dealers are plotted in Figures~\ref{clientgauss23456buybin} and \ref{clientgauss23456sellbin}.\\

\vspace{-8mm}

\vspace{-5mm}

We observe the same features as in the full-participation model. The values are just updated to account for the refined modelling of dealers' behavior.

\subsection{Best prices and hit ratios}

In this subsection we use the partial-participation model in order to answer some questions faced by practitioners.\\

For a dealer it is important to have an idea of the prices competitors will propose to the client during a specific RFQ. In fact it is even more important to have an estimation of the distribution of the best price that will be proposed to the client. Using the distributions of dealers' quotes $(F_0(\cdot;n))_n$ and $(F^*_0(\cdot;n))_n$ estimated in Section 4.2.2, it is straightforward to derive the distribution of the best quote proposed by competitors.\\

\begin{figure}[H]
  \centering
  \includegraphics[width=290pt]{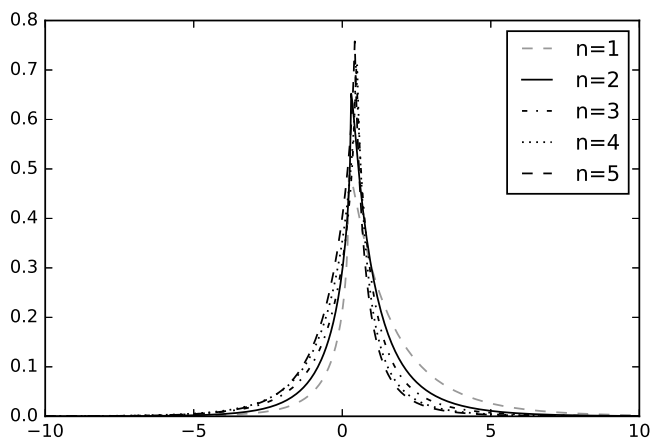}\\
  \caption{Conditional distribution of the best (reduced) quote proposed by competitors to clients, given that at least one dealer answers, calculated on each subsample of ``buy'' RFQs. Gray dashed line: $n=1$. Black solid line: $n=2$. Black dash-dotted line: $n=3$. Black dotted line: $n=4$.
  Black dashed line: $n=5$.}\label{bestpricebuybin}
\end{figure}

Let us start with the case of a ``buy'' order that will be answered by the reference dealer. If $n\geq 1$ dealers are requested during a RFQ in addition to the ``reference dealer'', there is a probability $(1 - p(n))^{n}$ that none of them answer. If we consider the more interesting case corresponding to at least one dealer answering the RFQ, then the (conditional) probability density function of the lowest (reduced) quote proposed by competitors is given by: $$\delta\mapsto \frac{1}{1- (1 - p(n))^{n} } f_0(\delta;n) \sum_{k=1}^{n} k \binom{n}{k} p(n)^{k} (1-p(n))^{n-k}  \left(1-F_0(\delta;n)\right)^{k-1}.$$
This probability density function is plotted in Figure~\ref{bestpricebuybin}.\\

\begin{figure}[H]
  \centering
  \includegraphics[width=290pt]{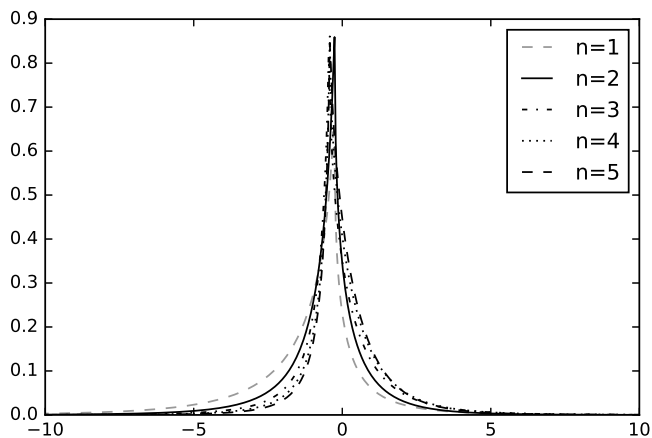}\\
  \caption{Conditional distribution of the best (reduced) quote proposed by competitors to clients, given that at least one dealer answers, calculated on each subsample of ``sell'' RFQs. Gray dashed line: $n=1$. Black solid line: $n=2$. Black dash-dotted line: $n=3$. Black dotted line: $n=4$.
  Black dashed line: $n=5$.}\label{bestpricesellbin}
\end{figure}

In the case of a ``sell'' RFQ, we obtain that the conditional probability density function of the highest price proposed by competitors when $n$ dealers are requested in addition the reference dealer, given that at least one dealer answers, is given by:
$$\delta\mapsto \frac{1}{1- (1 - p^*(n))^{n} } f_0^*(\delta;n) \sum_{k=1}^{n} k \binom{n}{k} p^*(n)^{k} (1-p^*(n))^{n-k}  F_0^*(\delta;n)^{k-1}.$$
This probability density function is plotted in Figure~\ref{bestpricesellbin}.\\

\vspace{-8mm}

What we learn from Figures \ref{bestpricebuybin} and \ref{bestpricesellbin} is that the skewness of the distribution of the best price evolves monotonically with the number of requested dealers.\footnote{Interestingly, the modes of the distributions in both Figure \ref{bestpricebuybin} and Figure \ref{bestpricesellbin} are very close across values of $n$.} In particular, competition plays its natural role: the larger the number of requested dealers, the lower (resp. higher) the price proposed by the best competitor in the case of a ``buy'' order (resp. ``sell'' order).\\

Moreover, corporate bond dealers are often interested in hit ratios. Hit ratios correspond to the probability to trade, given that the RFQ led to a trade with one of the competing dealers. In what follows, we show how the partial-participation model can be used to estimate (\emph{ex-ante}) hit ratio functions.\\

For a ``buy'' RFQ, given that $n$ dealers are requested in addition to the reference dealer, the probability to deal with the client by sending the reduced quote $\delta$, conditionally on the fact that there will be a deal with the client is:
\begin{eqnarray*}
\mathrm{HR}_{\text{buy}}(\delta) &=& P\left( V \ge \delta , \min_{k \leq \tilde n} W_{k} \ge \delta \, | \, V \geq \min(\delta,\min_{k\leq \tilde n} W_{k})    \right)  \\
&=& \frac{P\left( V \ge \delta , \min_{k \leq \tilde n} W_{k} \ge \delta  \right)}{1- P\left( V < \delta , V < \min_{k \leq \tilde n} W_{k}  \right)} \\
&=& \frac{\sum_{j=0}^{n} \binom{n}{k} p(n)^{k} (1-p(n))^{n-k} (1-G_0(\delta;n))(1-F_0(\delta;n))^k}{1 - \sum_{k=0}^{n} \binom{n}{k} p(n)^{k} (1-p(n))^{n-k} \int_{-\infty}^{\delta}  (1-F_0(v;n))^{k} g_0(v;n) dv}.
\end{eqnarray*}

Similarly, for a ``sell'' RFQ, given that $n$ dealers are requested in addition to the reference dealer, the probability to deal with the client by sending the reduced quote $\delta$, conditionally on the fact that there will be a deal with the client is:
\begin{eqnarray*}
  \mathrm{HR}_{\text{sell}}(\delta) &=& P\left( V \leq \delta , \max_{k \leq \tilde n} W_{k} \leq \delta \, | \, V \leq \max(\delta,\max_{k\leq \tilde n} W_{k})    \right)  \\
&=& \frac{P\left( V \leq \delta , \max_{k \leq \tilde n} W_{k} \leq \delta  \right)}{1- P\left( V > \delta , V > \max_{k \leq \tilde n} W_{k}  \right)} \\
&=& \frac{\sum_{k=0}^{n} \binom{n}{k} p(n)^{k} (1-p(n))^{n-k} G^*_0(\delta;n)F_0(\delta;n)^k}{1 - \sum_{k=0}^{n} \binom{n}{k} p(n)^{k} (1-p(n))^{n-k} \int^{+\infty}_{\delta}  F^*_0(v;n)^{k} g^*_0(v;n) dv}.
\end{eqnarray*}
Note that the denominator of the latter hit ratios is simply one minus the probability of getting a ``Not Traded'' RFQ.\\

\begin{figure}[H]
  \centering
  \includegraphics[width=290pt]{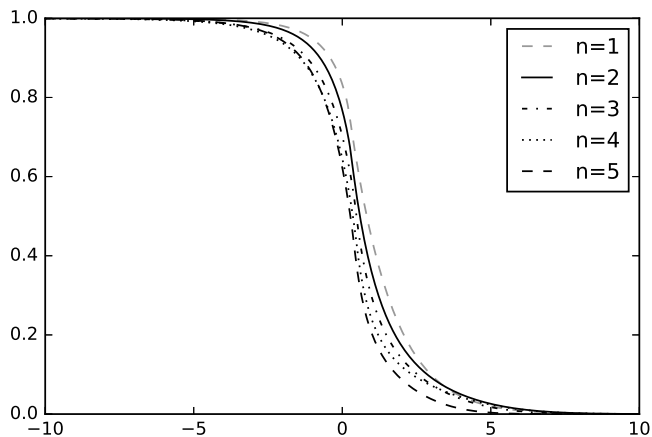}\\
  \caption{Hit ratios for a ``buy'' RFQ, as a function of the answered (reduced) quote, for each possible value of the number of competing dealers. Gray dashed line: $n=1$. Black solid line: $n=2$. Black dash-dotted line: $n=3$. Black dotted line: $n=4$.
  Black dashed line: $n=5$.}\label{proba_hr_buy_bin}
\end{figure}

\begin{figure}[H]
  \centering
  \includegraphics[width=290pt]{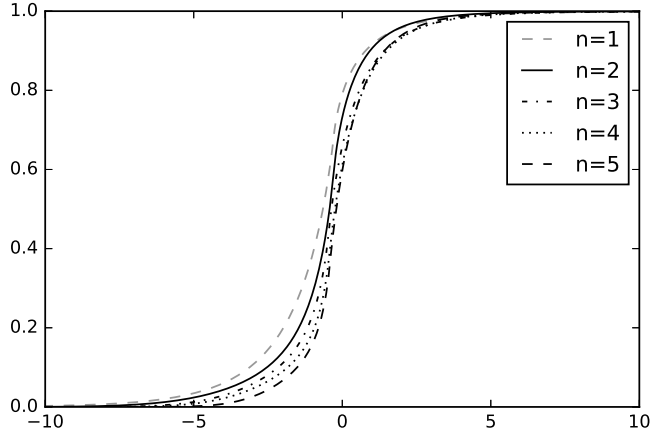}\\
  \caption{Hit ratios for a ``sell'' RFQ, as a function of the answered (reduced) quote, for each possible value of the number of competing dealers. Gray dashed line: $n=1$. Black solid line: $n=2$. Black dash-dotted line: $n=3$. Black dotted line: $n=4$.
  Black dashed line: $n=5$.}\label{proba_hr_sell_bin}
\end{figure}

These hit ratios are plotted in Figures~\ref{proba_hr_buy_bin} and~\ref{proba_hr_sell_bin}. Unsurprisingly, the hit ratios are monotonic functions of the price answered by the dealer: in the case of a ``buy'' (resp. ``sell'') RFQ, the probability to propose the best price decreases (resp. increases) with the price. Furthermore, hit ratios are decreasing functions of the number of requested dealers: the more competing dealers, the lower the probability to win the deal, for a given offered price.\\

\subsection{Model with covariates}

We now consider the introduction of various covariates into the partial-participation model. We consider, for each RFQ $i$, a vector $Z_i$ that concatenates extra-information about the bond characteristics, the client request, etc. $Z_i$ is assumed to be observed by the client and by all the requested dealers, \emph{i.e.}, $Z_i$ belongs to the RFQ information set $\Omega_i$.\\

Given $Z_i$, the law of dealers' quotes is assumed to be
\begin{equation}
 F( \xi|\Omega_i ) = F_0\left(\frac{\xi-\text{CBBT}_i}{\Delta_i}+ Z'_i\beta \right), \;\text{or}\; F^*( \xi|\Omega_i ) = F^*_0\left(\frac{\xi-\text{CBBT}_i}{\Delta_i} +Z_i'\beta^* \right),
\label{law_quotes_buy}
\end{equation}
for some cumulative distribution functions $F_0$ and $F^*_0$. In what follows, we consider SEP distributions for $F_0$ and $F^*_0$, as above.\footnote{It is noteworthy that the parameters of these SEP distributions have no reason to be the same as those obtained in the model without covariates.}\\

Similarly, given $Z_i$, the distribution of clients' quotes is
\begin{equation}
 G( \xi|\Omega_i ) = G_0\left(\frac{\xi-\text{CBBT}_i}{\Delta_i} + Z'_i\gamma  \right), \;\text{or}\; G^*( \xi|\Omega_i ) = G^*_0\left(\frac{\xi-\text{CBBT}_i}{\Delta_i} + Z'_i\gamma^* \right),
\label{law_quotes_sell}
\end{equation}
for some cumulative distribution functions $G_0$ and  $G^*_0$. As above, we consider Gaussian distributions for $G_0$ and  $G^*_0$.\\

In what follows we consider several extended models corresponding to different covariates. More precisely, we consider the following dummy variables:
\begin{itemize}
\item the requested bond is ``High-Yield'' (speculative grade),
\item its requested bond is ``Subordinated'', \emph{i.e.}, associated to a poor ranking in the firm capital structure,
\item in the RFQ, the requested notional is below  $100,000$ euros (this threshold corresponds to the first quartile of the bond notional distribution),
\item in the RFQ, the requested notional is above $500,000$ euros (this threshold corresponds to the third quartile of the bond notional distribution).
\end{itemize}

Instead of considering all these dummy variables at the same time, we consider three different extended models in order to reduce computation time. In the first one, we consider the marginal impact of the rating (whether or not the bond is a high-yield bond). In the second one, we consider the marginal impact of the seniority (whether or not the bond is a subordinated bond). In the third and last extended model, we consider the marginal impact of the bond notional by considering the two dummy variables corresponding to small and large notional respectively. The results of our estimation appear in Tables~\ref{res_all_covar1}, \ref{res_all_covar2}, and \ref{res_all_covar3}.\\

\begin{table}[H]
\begin{center}
{\footnotesize
\begin{tabular}{|c|c|c|c|c|c|c|c|c|c|} \hline
&\multicolumn{6}{|c|}{\text{Dealers}} & \multicolumn{3}{|c|}{\text{Clients}}  \\ \hline
\multirow{2}{*}{``Buy'' orders} &$\alpha$ & $\lambda$ & $\mu$ & $\sigma$ & $\beta$ (HY)  & $p$ & $\nu$ & $\tau$ & $\gamma$ (HY)   \\ \cline{2-10}
& $0.737$&  $0.262$&  $0.394$&  $0.883$ & $0.379$  & $0.388$ &  $1.96$& $2.00$ & $0.710$\\ \hline
\multirow{2}{*}{``Sell'' orders} &$\alpha^*$ & $\lambda^*$ & $\mu^*$ & $\sigma^*$ & $\beta^*$ (HY) & $p^*$ & $\nu^*$ & $\tau^*$ & $\gamma^*$ (HY)  \\ \cline{2-10}
& $0.665$& $-0.141$& $-0.403$&  $0.771$& $-0.267$ & $0.415$ &$-2.01$& $1.71$ & $-0.673$ \\ \hline
\end{tabular} \caption{Estimation of the first extended partial-participation model parameters, with the dummy variable ``High-Yield'', for all ``buy'' and ``sell'' orders.}\label{res_all_covar1} }
\end{center}
\end{table}

\begin{table}[H]
\begin{center}
{\footnotesize
\begin{tabular}{|c|c|c|c|c|c|c|c|c|c|} \hline
&\multicolumn{6}{|c|}{\text{Dealers}} & \multicolumn{3}{|c|}{\text{Clients}}  \\ \hline
\multirow{2}{*}{``Buy'' orders} &$\alpha$ & $\lambda$ & $\mu$ & $\sigma$ & $\beta$ (Sub.)  & $p$ & $\nu$ & $\tau$ & $\gamma$ (Sub.)   \\ \cline{2-10}
& $0.744$&  $0.305$&  $0.379$&  $0.880$ & $0.436$  & $0.385$ &  $1.91$& $1.91$ & $0.657$\\ \hline
\multirow{2}{*}{``Sell'' orders} &$\alpha^*$ & $\lambda^*$ & $\mu^*$ & $\sigma^*$ & $\beta^*$ (Sub.) & $p^*$ & $\nu^*$ & $\tau^*$ & $\gamma^*$ (Sub.)  \\ \cline{2-10}
& $0.679$& $-0.202$& $-0.358$&  $0.769$& $-0.326$ & $0.414$ &$-2.08$& $1.68$ & $-0.848$ \\ \hline
\end{tabular} \caption{Estimation of the second extended partial-participation model parameters, with the dummy variable ``Subordinated'', for all ``buy'' and ``sell'' orders.}\label{res_all_covar2} }
\end{center}
\end{table}

\begin{table}[H]
\begin{center}
{\footnotesize
\begin{tabular}{|c|c|c|c|c|c|c|c|} \hline
&\multicolumn{7}{|c|}{\text{Dealers}} \\ \hline
\multirow{2}{*}{``Buy'' orders} &$\alpha$ & $\lambda$ & $\mu$ & $\sigma$ & $\beta_{\text{Low}}$  & $\beta_{\text{High}}$ & $p$ \\ \cline{2-8}
& $0.733$&  $0.278$&  $0.384$&  $0.881$ & $0.478$ & $0.381$ & $0.386$ \\ \hline
\multirow{2}{*}{``Sell'' orders} &$\alpha^*$ & $\lambda^*$ & $\mu^*$ & $\sigma^*$ & $\beta^*_{\text{Low}}$ & $\beta^*_{\text{High}}$ & $p^*$ \\ \cline{2-8}
& $0.666$&  $-0.233$&  $-0.395$&  $0.746$ & $-0.630$ & $-0.488 $ & $0.404$ \\ \hline
\end{tabular}}
\end{center}
\end{table}

\vspace{-5mm}
\begin{table}[H]
\begin{center}
{\footnotesize
\begin{tabular}{|c|c|c|c|c|} \hline
& \multicolumn{4}{|c|}{\text{Clients}}  \\ \hline
\multirow{2}{*}{``Buy'' orders} & $\nu$ & $\tau$ & $\gamma_{\text{Low}}$ & $\gamma_{\text{High}}$ \\ \cline{2-5}
&   $1.83$& $2.01$ & $-0.323$ & $0.767$\\ \hline
\multirow{2}{*}{``Sell'' orders} & $\nu^*$ & $\tau^*$ & $\gamma^*_{\text{Low}}$ & $\gamma^*_{\text{High}}$ \\ \cline{2-5}
& $-1.88$& $1.84$ & $0.445$ & $-0.721$\\ \hline
\end{tabular}
\caption{Estimation of the second extended partial-participation model parameters, with the dummy variables ``Low Notional'' and ``High Notional'', for all ``buy'' and ``sell'' orders.}\label{res_all_covar3} }
\end{center}
\end{table}

In Tables~\ref{res_all_covar1}, \ref{res_all_covar2}, and \ref{res_all_covar3}, we see that the values of the parameters $(\alpha, \lambda, \mu, \sigma, p, \nu, \tau)$ and $(\alpha^*, \lambda^*, \mu^*, \sigma^*, p^*, \nu^*, \tau^*)$ are close to the values obtained before the addition of covariates. We also see the marginal impact of the considered covariates:

\begin{itemize}
  \item On average, when a client sends a RFQ, he/she is more demanding in terms of prices when the bond is High-Yield. To be specific, the mean client's reservation price  is decreased by $0.710$ times the CBBT bid-to-mid in the case of ``buy'' RFQs and increased by $0.673$ times the CBBT bid-to-mid in the case of ``sell'' RFQs . On the dealers side, requested dealers behave as if the CBBT was decreased by $0.379$ time the CBBT bid-to-mid in the case of ``buy'' RFQs, and increased by $0.267$ time the CBBT bid-to-mid in the case of ``sell'' RFQs. This discount probably reflects the higher risk aversion of clients with respect to bonds for which default events are not unlikely. Dealers behave accordingly, but the discount associated with speculative bonds is smaller on the dealers side than on the clients side.
  \item Very similar features are observed with subordinated bonds. In that case, the higher risk aversion is related to lower and more uncertain recovery rates in the case of a default event.
  \item As far as the notional is concerned, there are several effects at stake. When the notional of a RFQ is large, clients are looking for good prices. The dealers answering the request accept to propose interesting prices to the clients (although they do not accept the average discount expected on average by clients). In our partial-participation model, this effect may be due to the fact that the dealers answering the RFQ are those who have an inventory compatible with the request. In other words, the transaction (if it occurs) may reduce the total risk of their portfolio. Another possible rationale is that large bond buyers/sellers are probably important/strategic clients for most dealers, and they surely have the opportunity to obtain more interesting prices than other clients. For small notional RFQs, both dealers and clients seem to be ready (on average) to shift their quotes in order to facilitate transactions (compared to what happens for RFQs with average notional). Because small trades are less risky, this effect may be due to the willingness of dealers to increase the number of deals.\\
\end{itemize}

\section*{Conclusion}

In this paper, we have introduced a new modelling framework to infer the behavior of dealers and clients on the corporate bond market from a hitherto unexploited\footnote{Although practitioners have been using this kind of dataset for several years, it is the first time an academic research work is based on such a type of dataset.} database of RFQs.\\

We have modelled in a simple but realistic way the RFQ process on MD2C platforms by assuming skewed exponential power distributions for dealers' quotes and Gaussian distributions for clients' reservation prices. We have stressed the importance of the no-answer effect, and have proposed a way of measuring it. In our ``partial-participation'' model, we find that the distributions of dealers' quotes are more realistic than in the ``full-participation'' model, and depend only slightly on the number of requested dealers $n_i$. When this number increases, the probability of not answering increases on one hand, and the dealer quotes are more aggressive on the other hand. Globally and fortunately, in line with intuition, the level of competition among effective participants increases with $n_i$. This can be checked by the better prices that are offered to clients, and by the lower hit ratios that can be achieved by dealers when $n_i$ increases.\\

Our modeling framework could be improved in at least two ways. First, it would be possible to introduce other explanatory variables in the model: variables that are related to the market environment (traded volumes, stock or interest rate market volatility, news, etc.), the time of the RFQ (to account for intra-day seasonality), more bond-related information (identity of the issuer, recent price trends, etc.). Second, it would be very interesting to model the exact timing of the answers received by the client, the associated quotes, and the resulting client's behavior. The timing of the process stays largely unknown for now.

\vspace{3cm}

\begin{center}
\textbf{Acknowledgements}
\end{center}

The authors would like to thank Philippe Amzelek, Joe Bonnaud, Laurent Carlier, Jean-Michel Lasry, Andrei Serjantov, and Vladimir Vasiliev for the discussions they had on the subject. An anonymous referee also needs to be warmly thanked for his thorough reading of our paper. St\'ephane Gaiffas,  Arnaud Rachez, and Robin Ryder also need to be thanked for their contributions to a preliminary version of this work.

\bibliographystyle{plain}

\end{document}